\documentclass[sn-mathphys,Numbered]{sn-jnl}

\pdfoutput=1

\usepackage{multirow}%
\usepackage{mathrsfs}%
\usepackage[title]{appendix}%
\usepackage{xcolor}%
\usepackage{textcomp}%
\usepackage{manyfoot}%
\usepackage{booktabs}%
\usepackage{algorithm}%
\usepackage{algorithmicx}%
\usepackage{algpseudocode}%

\usepackage{amsmath,amssymb,bm,amsfonts}
\usepackage{fancyhdr,graphicx,pict2e}
\usepackage[greek,english]{babel}
\usepackage{natbib}
\usepackage{anysize}
\usepackage{setspace}
\usepackage[mathlines]{lineno}
\usepackage{amsthm}
\usepackage{tikz}
\usepackage{tikz-3dplot}
\usepackage{mathtools}
\usepackage{listings}
\usepackage{color} 
\definecolor{mygreen}{RGB}{28,172,0} 
\definecolor{mylilas}{RGB}{170,55,241}
\marginsize{2cm}{2cm}{2cm}{2cm}
\definecolor{lyell}{cmyk}{0,0,0.2,0}
\newcommand{\cob}{\color{blue}}
\definecolor{dred}{rgb}{0.85,0,0}
\newcommand{\cor}{\color{dred}}
\definecolor{dgreen}{rgb}{0,0.4,0}

\definecolor{dbrown}{cmyk}{0,0.30,0.45,0.64}

\definecolor{redy}{rgb}{0.7,0.6,0.6}

\definecolor{redb}{rgb}{0.7,0,0.7}

\definecolor{light}{gray}{.75}
\definecolor{dark}{gray}{.5}

\newcommand{\bc}{\begin{center}}
\newcommand{\ec}{\end{center}}
\newcommand{\bpic}{\begin{picture}}
\newcommand{\epic}{\end{picture}}
\newcommand{\ba}{\begin{array}}
\newcommand{\ea}{\end{array}}
\newcommand{\be}{\begin{eqnarray}}
\newcommand{\ee}{\end{eqnarray}}
\newcommand{\beq}{\begin{eqnarray*}}
\newcommand{\eeq}{\end{eqnarray*}}

\newcommand{\hsp}{\hspace*{\fill}}

\renewcommand{\d}{\mathrm{d}}
\newcommand{\D}{\mathrm{D}}
\newcommand{\e}{\mathrm{e}}

\newcommand{\lra}{\longrightarrow}

\newcommand{\dee}{\partial}
\newcommand{\del}{\boldsymbol{\nabla}}

\newcommand{\mb}[1]{\bm{#1}}

\newcommand{\R}[1]{\mathbb{R}^{#1}}
\newcommand{\oiint}{\bigcirc\hspace*{-1.42em}\int\hspace*{-0.8em}\int}
\newcommand{\sqs}{\sqrt{s}}

\renewcommand{\pi}{\textrm{\greektext p}}
\renewcommand{\theequation}{\arabic{equation}}

\begin{document}
\title[Particle uptake by a cell]{\center Nanoparticle uptake by a semi-permeable, spherical cell from an external planar diffusive field. I. Mathematical model and asymptotic solution.}
\author[1]{\fnm{Stanley J.} \sur{Miklavcic}}\email{stan.miklavcic@unisa.edu.au}
\affil[1]{\orgdiv{Phenomics and Bioinformatics Research Centre, UniSA STEM}, \orgname{University of South Australia}, \orgaddress{\city{Adelaide}, \postcode{5095}, \state{SA}, \country{Australia}}}
\abstract{In this paper we consider the diffusion of nanoparticles taken up by a semi-permeable spherical cell placed in the path of a diffusive particle field generated by an external planar source. The cell interior and exterior are characterized by different diffusive properties, while the cell is able to accommodate a different saturation level of particles at steady state than is present in the external medium. The situation models the practical problem of biological cells exposed from one direction. The conflict of geometries is handled by the introduction of an effective boundary condition at a virtual spherical boundary. A closed-form, large-time asymptotic solution for the local concentration interior to the cell is developed. We consequently derive an asymptotic approximation for the rate of nanoparticle accumulation in the cell. We contrast the resulting time dependence with that of the corresponding quantity found under strictly spherically symmetric conditions.}
\maketitle
\section{Introduction} \label{sec:introduction}

Common to both animal and plant physiology, living cells take up elements from external sources for their continued existence, development and function. Although a cell will have become adapted to optimally take up favorable elements (nutrients) to which it has been provided, in certain unfavorable circumstances it may also take up elements or compounds that could be detrimental to a cell's function, development, and indeed existence. In such circumstances one may hope that cells have also evolved to either prevent the uptake of toxic elements or to minimize the adverse effects of any toxic element that it has taken up. An example of this can be found with plant cells which, on the one hand, have developed strategies to prevent the uptake of toxic salt species (salt exclusion) or, on the other, have the possibility of loading toxic ions into their vacuoles, out of harms way (salt tolerance) \cite{Munns2008}. A more recent situation, to which cell evolution may not have caught up is exposure to nanoparticles. Recent literature interest in the subject of nanoparticle uptake in biological systems, is divided between the positive use of nanoparticles for biomedical applications such as medicinal treatments (\textit{e.g.}, therapeutics)~\cite{AlObaidi2015}, and the potentially harmful effects of nanoparticle accumulation in cells and tissues~\cite{Gogosti2003,Colvin2003}.

To quantify the benefits of, or harm by, nanoparticles, it is necessary to determine the amount of particle material that is taken up. Both the total amount and the rate of uptake are quantities characterizing a cell's propensity for nanoparticles. Since solute material is delivered to cells by means of diffusion, convection or by the combined effect of convective-diffusion~\cite{islam2017}, characterizing the diffusive and convective properties of the external medium is as important as characterizing the cell membrane's affinity for nanoparticles. Some not so recent works \cite{rashevsky1948,crank1956,philip1964,mild1971}\footnote{The paper by Philip~\cite{philip1964} explicitly modelled heat conduction, but the analysis is identical to that of particle diffusion.} reported on results of theoretical studies modelling solute transport internal or external to a spherical cell by the process of diffusion. More recent efforts expand on these early papers. For instance, Sorrell et al. \cite{Sorrell2014} and West et al. \cite{West2021} characterized particle internalization by means of a multi-event, mass action kinetic model which ignored the spatial extent of cells, yet included an array of membrane binding, reaction and transfer processes. This spatially independent model was complemented in West et al. \cite{West2021} by a spatial model of a spheroidal aggregate of cells. The aggregate properties were differentiated discretely in the radial direction but were otherwise spherically symmetric (and homogeneous). A similarly structured, spherically symmetric model was employed by Foster and Miklavcic \cite{foster2015} to study salt uptake by a plant cell. The degree of biological and chemical detail included in these studies has varied, as have their aims, yet each has provided insight into some aspect of solute movement. What is common to most models (apart from an underlying assumption of the diffusion process) is the simplifying assumption of spherical symmetry. This of course limits the applicability of these models to highly specialized experimental and practical situations \cite{Zanoni2016}. Recent experimental studies of nanoparticle uptake, conducted by the authors of \cite{Limbach2005} and \cite{Cui2016}, on the other hand, feature cells exposed unilaterally, not isotropically, to nanoparticle solutions: cells were seeded and allowed to adhere to planar glass substrates, which were then exposed to nanoparticle solutions. Among other findings, such as particle concentration dependence, particle uptake was found to depend on nanoparticle size \emph{and} the orientation of the cell-covered substrates, the latter specifically indicating a role played by sedimentation on uptake rate.

The experimental studies and findings are clearly inconsistent with the classical model of a spherical cell exposed isotropically to a particulate solution. This simplifying assumption, while analytically expedient, leads to a specific prediction of particle uptake rate and dependence on cell properties that fails to recognize the actual anisotropic exposure and which cannot accommodate particle sedimentation. In this same connection we mention that to entertain a possible theoretical comparison with experiments, the authors of \cite{Limbach2005} and \cite{Cui2016} adopted a one-dimensional model of diffusion with sedimentation. Unfortunately, a unidirectional diffusion and sedimentation model offers little scope for a rigorous study of the time-dependent diffusive behaviour in and around an individual cell. Specifically, such a simplified approach cannot provide an accurate account of the dependence of the rate of particle uptake on cell properties.

Works detailing particle diffusion that depart from spherical symmetry are few \cite{AlObaidi2015,Hinderliter2010,Friedmann2016} and largely rely on 3D numerical simulations. Relevant exceptions include a Green function approach for the point-to-point diffusion within a spherical cell with an absorbing surface \cite{Arjmandi2019} and the works by Sch\"{a}fer and coworkers \cite{Schafer2019b,Schafer2020} who treated diffusion within a spherical cell with a semi-permeable boundary to which is applied general boundary behavior. The latter work built on (transfer function) methods developed for other anisotropic situations \cite{Schafer2018,Schafer2019a}. What concerns us in this paper is the construction and solution of a mathematical model describing the unidirectional diffusion of particles toward, and subsequent uptake by, a spherical cell. On the one hand the present work may be thought of as a special case of the general system imagined by Sch\"{a}fer et al. \cite{Schafer2019b,Schafer2020}. On the other hand, the problem here is more intricate in that solution to the diffusion problem in the external medium is simultaneously also required. By resolving explicitly the conflict of geometries arising in this diffusion problem we can remove the confounding effect of unidirectional contra isotropic exposure. In the process, we expose the true dependence of particle uptake on the cell's intrinsic membrane properties. Such a mathematical model and its complete analytical solution has not been successfully attempted previously, to the author's knowledge. In this work we not only extend previous diffusion models in this direction, we also establish how more details of a biological nature may be incorporated. Some aspects of the manner of solution, as well as the solution itself, are also of notable interest. The analytic results, in the form of asymptotic series, have been compared with numerical (FEM) solutions and found to be highly accurate under a wide range of circumstances. For reasons of space a numerical study of accuracy is presented in a follow-up paper~\cite{Kumar2024}.

In Section \ref{sec:method} we present the physical model and the relevant governing equations neglecting sedimentation for the present (Appendix A describes one approach toward including sedimentation). The governing equations are then solved in Section \ref{sec:solution} to the analytical extent possible using the relevant methods of Laplace transform and separation of variables. The fundamental approach is equivalent to the transfer function formulation of Sch\"{a}fer et al. \cite{Schafer2019b,Schafer2020}, but goes further with the development of explicit variable and parameter dependent solutions. Only a summary of the key results are given in Section \ref{sec:solution}. Important detailed information relevant to the solution process has been relegated to Appendices B-E. Briefly, in Section \ref{sec:solution} the penultimate objective quantity of interest, the local, time-dependent solute concentration \emph{interior} to the cell, is presented in the form of an eigenfunction expansion with coefficients that are expressed as asymptotic series in the time variable. Various solution approximations may be obtained by appropriate truncation of these series. In Appendix \ref{sec:largetimesolution} we explore a few of the more significant of these approximations. To highlight the significance of our main contribution we provide, for comparison, in Section \ref{sec:sphericalcase} a parallel study of nanoparticle diffusion in the spherically symmetric system most closely associated with the anisotropic system: a spherical cell centered within a spherical domain whose outer boundary is a uniform source of diffusing particles. Section \ref{sec:discussion1} contains the key result of this paper. The more robust of the approximations considered in Appendix \ref{sec:largetimesolution} is applied in Section \ref{sec:discussion1} to determine the time-dependent accumulation of nanoparticles taken up by the cell. This quantity is arguably of greater experimental interest than is the local solute concentration itself. The closed expression we derive is in striking contrast with the corresponding result for the spherically symmetric system, a contrast which simply underscores the significance of the mode of delivery over and above the cell's intrinsic properties. We summarize our findings in Section \ref{sec:conclusion} and provide some suggestions for future work.
\begin{figure}[t!]
    \centering
 \begin{tikzpicture}[scale=0.8,  rotate=90]
   \def\w{8}
   \draw[black,line width=1 mm,dashed] (5.15,-\w) -- (5.15,\w);
   \draw[black,line width=0.5mm,dashed] (4.25,-\w) -- (4.25,\w);
   \draw[black,line width=0.25mm,dashed] (3.5,-\w) -- (3.5,\w);
   \draw[black,line width=0.125mm,dashed] (2.75,-\w) -- (2.75,\w);
   \fill[draw=black,fill=black!10,line width=0.5mm,dashed] (0,0) circle (1.8cm);
    \draw[black, dotted, line width=0.25mm] (0,0) circle (4.75 cm);
    \draw[black, dashed, line width=0.5mm] (0,0) circle (2.0 cm);
   \fill[black!10] (5.2,\w) rectangle (8.5,-\w);
   \node[align=center] at (-0.35,0.5) {Cell~~~\\ (Region $II$)\\$c_{II},\,D_{II}$};
   \draw[-,draw=black,line width=0.35mm] (0,0) -- (5.18,0) node[at end, left, xshift=-0.1cm,yshift=0.25cm] {$z=z_0$} ;
   \draw[-,draw=black] (0,0) -- (1.4,-1.4) node[at end, pos=1.2] {$a$} ;
   \draw[-,draw=black] (0,0) -- (-0.6,-1.7) node[below right, pos=1.2] {$a-v$} ;
    \draw[->] (0,0) --(8,0) node[right] {$z$} ;
   \draw[->] (0,0) --(0,-7) node[right] {$x$} ;
    \draw[-,draw=black] (0,0) -- (4.385,-1.827) node[below right, pos=0.7, xshift=0.2cm,yshift=0.4cm] {$R$} ;
   \node[align=center] at (1.25,4.5) { External Medium \\ (Region $I$) \\ $c_{I},\,D_{I}$};
   \node[align=left] at (7,-3) {Particle\\~~~~ Reservoir\\~~~~~~~~~~$C_0$};
   \draw[|->,black, line width=0.35mm] (5.15,-7) -- (3.75,-7) node[right, pos=0.85, xshift=  0.25cm] {$\mathbf{N_{surf}}$};
   \def\a{2}
   \coordinate (circleCenter) at (-1,1);
        \coordinate (pointOnCircle) at (-1.414,1.414);
        \draw[|->,black, line width=0.35mm] (pointOnCircle) -- ($(pointOnCircle)!-\a!(circleCenter)$) node[above, pos=1.0, xshift=  -0.2cm] {$\mathbf{N}$};
\end{tikzpicture}
   \caption{Schematic of the diffusion problem of a spherical cell (dashed circles) of outer radius $a$ and membrane thickness $\upsilon$ in the path of a diffusing front of particles emerging from an infinite half-space reservoir source (shaded rectangle) bounded below at $z=z_0$. The particles diffuse uniformly in the direction ($\mathbf{N}_{surf}=(0,0,-1)$) into the semi-infinite Region I ($z<z_0$) as depicted by the horizontal lines of graded thickness which represent level-set particle concentrations.  The cell's outward pointing unit vector normal is $\mathbf{N}=\widehat{\mathbf{r}}$. Region $I$ is characterized by diffusion constant $D_I$, while in Region $II$ the diffusion constant is $D_{II}$. The large dotted circle of radius $R$ shows the location where the effective far-field boundary condition Eq \eqref{OuterBC} is applied. The schematic is not drawn to scale to ensure clarity of definitions. The cell is centered at the origin of a Cartesian coordinate system ($z$-axis directed upward). \label{fig:figure1}}
\end{figure}
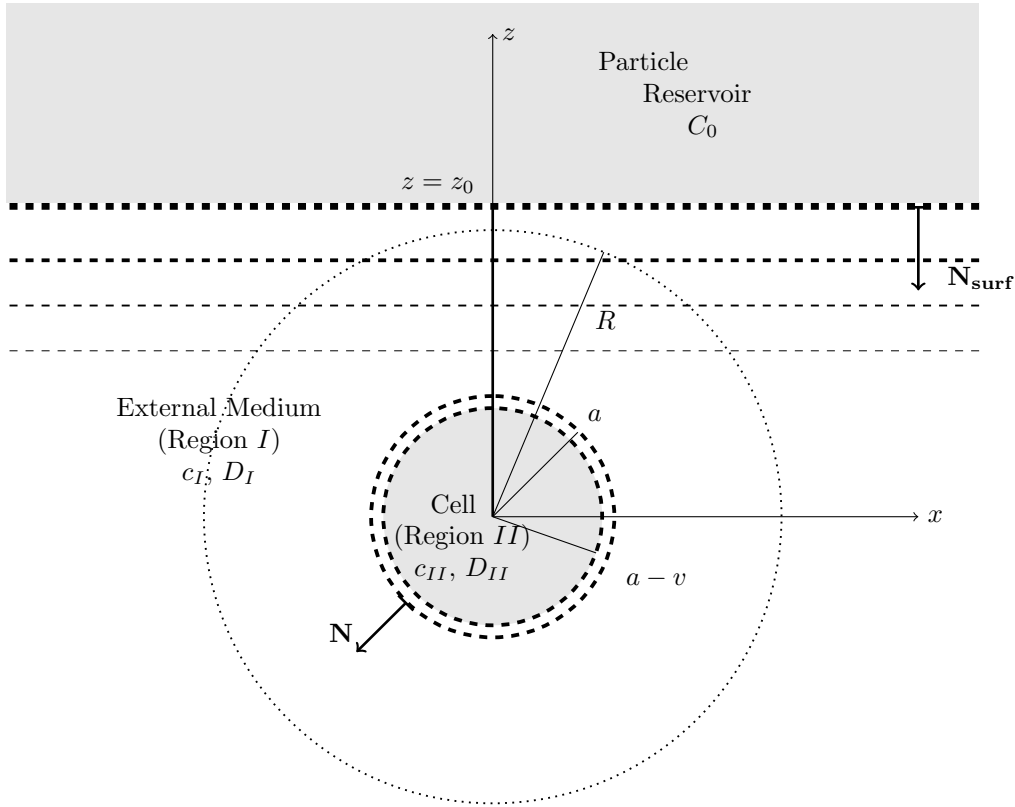

\section{A Spherical Cell in a Unidirectional Particle Field} \label{sec:method}
\subsection{General Comments on the Physical and Mathematical Model}
We assume the cell in question is spherical and comprises only one membrane surface which separates the inside from the outside. The neutral (uncharged) solutes to which the cell is exposed approach the cell from one direction (from above, say). The scenario is shown schematically (not to scale) in Figure \ref{fig:figure1}. Since the diffusion-only problem involving conflicting geometries is already sufficiently complicated we shall explicitly treat the conjunction of diffusion and sedimentation in a separate paper and confine attention here to the diffusion-only problem. For the present we comment, firstly, that sedimentation does not appear to be an important factor for particle sizes of the order 25-50 nm~\cite{Limbach2005,Cui2016}, so the present analysis may be considered valid under those conditions. Secondly, sedimentation could be incorporated adhoc in the solution of the strictly diffusion-only problem by a suitable transformation, as outlined in Appendix A.

\subsection{Details of the Mathematical Model}

\smallskip

\subsubsection{Geometric Structure}
The model system comprises a spherical cell of outer radius, $a$ and inner radius, $a-\upsilon$. Internal to the cell, medium II, is a fluid environment of characteristic diffusivity coefficient $D_{II}$ (see Figure \ref{fig:figure1}). The medium external to the cell, medium I, is a fluid characterized by a diffusivity coefficient of $D_{I}$. For convenience, the sphere is positioned at the origin of a 3D coordinate system. In terms of Cartesian coordinates, the sphere center is then at $(0,0,0)$. Far from the sphere, at a distance $z_0>0$ from the origin, is a source of solutes/particles. If the source is macroscopic and sufficiently far from the cell, it may be assumed that the particles originate from a planar source which, from the perspective of a cell of relatively small radius, is a reasonable approximation.  The plane surface has a normal vector, $\mb{N}_{surf}=(0,0,-1)$, directed into medium I. We presume the source to be the boundary of an infinite reservoir of solute particles at a concentration $C_0$, which subsequently diffuse from the planar surface of the reservoir in the direction $\mb{N}_{surf}$ and thus in the direction of the spherical cell. We denote the local concentration of solute particles at time $t$ by $c_I(\mathbf{r},t)$ in the exterior medium $I$, between the particle reservoir and the cell, and by $c_{II}(\mathbf{r},t)$ in the medium $II$ interior to the cell. The geometric configuration of the plane and spherical cell is shown in Figure \ref{fig:figure1}.

\smallskip

\subsubsection{Governing Equations} \label{sec:governingeqns}
Although it is quite reasonable to suppose that the diffusive properties of the media inside and outside a cell are not ideally Newtonian, to keep the complications to a minimum we shall assume that both internal and external fluids can be described by Stokes-Einstein diffusion constants, $D_{I}$ and $D_{II}$, and the particles are sufficiently mobile to undertake random molecular motions thus governed by classical diffusion equations in the two respective regions:

\begin{eqnarray}
\dfrac{\partial c_{I}}{\partial t} &=& D_{I}\nabla^2c_{I} = D_{I}\left[\dfrac{1}{r^2}\dfrac{\dee}{\dee r}\left(r^2\dfrac{\dee c_{I}}{\dee r}\right)
+\dfrac{1}{r^2 \sin \phi}\dfrac{\dee}{\dee \phi}\left(\sin \phi \dfrac{\dee c_{I}}{\dee \phi}\right)
+\dfrac{1}{r^2 \sin^2 \phi} \dfrac{\dee c_{I}}{\dee \theta^2}\right], \label{diffRegI} \vspace{0.85cm}\\
&& \hspace{6.5cm} r>a, \quad \phi\in[0,\pi], \quad \theta \in [0,2\pi], \quad t>0, \nonumber \\
 \nonumber \\
\dfrac{\partial c_{II}}{\partial t} &=& D_{II}\nabla^2c_{II} = D_{II}\left[\dfrac{1}{r^2}\dfrac{\dee}{\dee r}\left(r^2\dfrac{\dee c_{II}}{\dee r}\right)
+\dfrac{1}{r^2 \sin \phi}\dfrac{\dee}{\dee \phi}\left(\sin \phi \dfrac{\dee c_{II}}{\dee \phi}\right)
+\dfrac{1}{r^2 \sin^2 \phi} \dfrac{\dee c_{II}}{\dee \theta^2}\right], \label{diffRegII} \vspace{0.85cm}\\
&& \hspace{6.5cm} 0<r<a-\upsilon, \quad \phi\in[0,\pi], \quad \theta \in [0,2\pi], \quad t>0. \nonumber
\end{eqnarray}

\smallskip

\subsubsection{Boundary and Initial Conditions} \label{sec:bandiconds}
To complement the above governing equations we need to adopt conditions that will ensure a unique solution. However, in doing so we attempt to include some flexibility in order to explore different physical states.

To begin with we specify our initial condition that prior to time $t=0$, the region inside the cell and the infinite half space, $z < z_0$, outside the cell is devoid of solutes,
\begin{equation}
\left\{
\begin{array}{l}
c_{I}(\cdot, t = 0) = 0, \quad \quad \text{ \ in region I },\\
c_{II}(\cdot, t = 0) = 0, \quad \quad \text{ in region II },
\end{array}
\right. \label{icond}
\end{equation}

The first boundary condition relates the normal components of the solute surface fluxes at the inside and outside cell walls:
\begin{equation*}
\left.\left[-\mb{N} \cdot  \left(-D_{II} \del c_{II}\right) \right]\right|_{(a-\upsilon,\phi, t)} = \alpha \left.\left[-\mb{N} \cdot \left(-D_{I}\del c_{I}\right) \right]\right|_{(a,\phi,t)}, \quad \phi\in[0,\pi],~t > 0,
\end{equation*}
which, given the spherical cell geometry with unit normal vector $\mb{N} = \widehat{\mb{r}}$, reduces to
\begin{equation}
\left.\left[-D_{II}\left(\dfrac{\partial c_{II}}{\partial r}\right)\right]\right|_{(a-\upsilon,\phi, t)} = \alpha \left.\left[-D_{I}\left(\dfrac{\partial c_{I}}{\partial r}\right)\right]\right|_{(a,\phi,t)}, \quad \phi\in[0,\pi],~t > 0, \label{cond:flux}
\end{equation}
where $0 < \alpha \le 1$. The second boundary condition relates the concentration of solutes just inside the cell to that just outside:
\begin{equation}
c_{II}(r=a-\upsilon,\phi,t) = \beta c_{I}(r=a,\phi,t), \quad \quad \phi\in[0,\pi],~t > 0, \hsp
\label{cond:conc}
\end{equation}
where $\beta \ge 0$. We remark that since the diffusing particles are uncharged, no consideration of membrane potential effects are included here.

Our final boundary conditions reflect, on the one hand, the continual provision of solutes at the plane surface of the reservoir and, on the other, the fact that far from both the sphere and the plane source, the concentration of solutes approaches zero:
\begin{multline}
\left\{
\begin{array}{l}
c_{I}(\mb{r},t) = C_0, \hspace{4.5cm} z=z_0, \quad x,y \in \R{2}, \quad t>0,\\
c_{I}(\mb{r},t) \lra 0, \hspace{4.4cm} z \rightarrow -\infty, \quad x,y \in \R{2}, \quad t>0.
\end{array}
\right. \hsp \label{plnbc}
\end{multline}

\subsubsection{Comments on Parameters}

Actual biological cells are complex structures. The exterior of a plant cell, as a particular example, comprises a finitely thick region of cellulose microfibrils embedded in a matrix of proteins and other macromolecules. Internal to this so-called \textit{cell wall} lies the cell's phospholipid bilayer membrane, \textit{the plasma membrane}, which contains contains hydrophobic proteins and also supports hydrophilic proteins. Occupying the greater part of the cell interior (approximately 80-90\% of a mature cell's volume) is the cell's \textit{vacuole} which is bounded by \textit{the tonoplast membrane}. Finally, between the two membranes is the cell's \textit{cytoplasm}~\citep{bidwell1979,salisbury1991}.

To advocate the use of our idealized model to represent such a complex entity we have introduced three parameters, $\alpha$, $\beta$, and $\upsilon$, to which we might assign phenomenological roles. As in a previous effort~\citep{foster2015} our interest focuses on the vacuole as the dominant cell storage compartment, and the external medium being that which exists beyond the cell wall, the parameter $\upsilon$ thus effectively accounts not only for the finite thicknesses of the cell wall, the cytoplasm and the two membranes, to some degree it is a measure of particle accumulation within these regions. We use $\beta$ in \eqref{cond:conc} to allow for the very real possibility that at steady state the concentration of solutes inside the cell vacuole is significantly different to, either greater than or less than, the steady state concentration outside the cell~\citep{crank1956}. This is physically possible in plant cells, where either toxic molecules are passively and actively sequestered into the cell's vacuole out of harm's way ($\beta > 1$), allowing the cell to continue functioning. Alternatively, there may be active mechanisms preventing the passive build up of particles ($0 < \beta < 1$)~\citep{bidwell1979,salisbury1991}. If $\left[-\mb{N} \cdot \left(-D_{I} \del c_{I}(a,t)\right)\right]$ represents the normal flux per unit area of solutes arriving at the outer cell boundary~\citep{rashevsky1948}, then with $\alpha <1$ in condition \eqref{cond:flux} we may account for the possibility of a reduced particle flux entering the cell vacuole, being some fraction of that arriving at the outer boundary. In other words, not all solutes arriving in a direction normal to the cell will enter the cell. Thus, $\alpha$ can be used as a measure of a cell's propensity to take up (through absorption or adsorption) of particles by the cell wall, the plasma or tonoplast membranes, or the cytoplasm, reducing the rate of uptake by the vacuole. Note that we explicitly restrict $\alpha \le 1$ as $\alpha$ greater than unity would suggest that more solutes enter the cell than actually arrive, which is physically impossible.

While the roles played by $\alpha$, $\beta$, and $\upsilon$ may be unfamiliar we point out that including these parameters requires very little additional effort, but is beneficial in the sense that with their inclusion we may discover where and with what importance they feature in the end result. Nevertheless, the more conservative reader may be reassured that by setting the two parameter constants, $\alpha$ and $\beta$, to unity, and membrane thickness $\upsilon$ to zero, the boundary conditions reduce to the textbook boundary conditions entertained in \cite{philip1964} and \cite{mild1971}, and others. That is, with $\alpha = \beta = 1$ and with $\upsilon=0$, conditions \eqref{cond:flux} and \eqref{cond:conc} reduce simply to statements of the usual conditions of continuity of concentration and continuity of normal flux, respectively, which are traditionally applied to a mathematically thin, idealized membrane~\citep{crank1956}. This is in fact the case treated in Paper II by Kumar and Miklavcic \cite{Kumar2024} in their comparison of the numerical implementation of the asymptotic solution with a fully numerical, finite element method solution.

\bigskip

\section{Solution to the Anisotropic Diffusion Problem} \label{sec:solution}

\subsection{Re-specification of the Far-field Boundary Condition} \label{sec:farfield}
The physical problem as has been introduced thus far, and particularly the pair of conditions in  \eqref{plnbc}, clearly introduces a complication due to a conflict of geometries. On the one hand there is the planar source of particles, at $z=z_0$, which induces a plane diffusive front of particles moving on average in the negative $z$ direction. On the other hand, there is the spherical cell of radius $a$ at the origin, which meets this front. The combination of the two aspects actually results in an axisymmetric system, with no dependence on azimuthal angle, only a dependence on the radial and polar angle coordinates. This reduction in variable dependence does not, unfortunately, diminish the complication arising from the application of boundary condition \eqref{plnbc} at the plane source, for all $x$ and $y$ at the same time as satisfying the zero concentration limit condition as $z \rightarrow -\infty$, for all $x$ and $y$. However, we may simplify the problem by invoking the following construct.

In the case where there is no spherical cell, the plane source of particles produces a diffusion profile that is spatially one-dimensional, depending only on $z$ (or, actually $z-z_0$). This problem is readily described by the time-dependent diffusion problem in one spatial dimension:
\begin{multline}
\left\{
\begin{array}{l}
\dfrac{\dee c_{\infty}}{\dee t} = D_I\dfrac{\dee^2 c_{\infty}}{\dee z^2}, \hspace{4cm} \text{for} \quad z < z_0, \quad t > 0, \hsp\\
c_{\infty}(z,t) = C_0, \hspace{4.5cm} \text{for} \quad z=z_0, \quad t \ge0, \hsp\\
c_{\infty}(z,t) \lra 0, \hspace{4.4cm} \text{as} \quad z \lra -\infty, \quad t \ge 0. \hsp
\end{array}
\right. \label{OneDprob}
\hsp
\end{multline}
The problem has a well known solution~\citep{crank1956}, obtained by means of the Laplace transform method to be described below. In the present context this solution takes the form,
\begin{equation}
c_{\infty}(z,t) = C_0 \left(1 - \text{erf}\left(\dfrac{(z_0-z)}{\sqrt{4 D_I t}}\right)\right), \quad \quad -\infty < z \le z_0, \quad t \ge 0, \hsp \label{OneDsoln}
\end{equation}
where $\text{erf}(y) = \frac{2}{\sqrt{\pi}}\int_0^y \e^{-x^2} \d x$ is the error function~\citep{abram1965}. Obviously, the level surfaces of $c_{\infty}$ are planes parallel to $z=z_0$ (as depicted schematically in Figure \ref{fig:figure1}). On the other hand, on a virtual spherical surface, of radius $R \le z_0$ and centred at the origin (represented by the large outer circle in Figure \ref{fig:figure1}), the solute concentration at any instant of time will be given by $c_{\infty}(z=R\cos \phi,t)$, where $\phi \in [0,\pi]$.

We now suppose we have the physical, spherical cell of radius $a$ centered at the origin. In its presence, provided $a \ll R \le z_0$, the diffusive field described by  \eqref{OneDsoln} will, to a very good approximation, remain unperturbed very far from the cell. In particular, this state will be true at the position of the virtual spherical surface. This physical fact may be utilized to replace the pair of boundary conditions in  \eqref{plnbc} with the following, inhomogeneous spherical boundary condition at $r=R$,
\begin{equation}
c_{I}(R,\phi,t) = c_{\infty}(R \cos \phi,t),  \quad \quad \quad r = R, \quad \phi \in[0,\pi], \quad t > 0. \hsp \label{OuterBC}
\end{equation}
We note that if diffusion within the cell is less rapid than diffusion outside the cell (\textit{i.e.}, $D_{II} < D_{I}$) then the external diffusive field far from the cell will be even less affected by the cell's presence, and the approximation becomes even better.

In our problem, instead of  \eqref{plnbc} we invoke  \eqref{OuterBC}, which automatically satisfies both physical boundary conditions specified in  \eqref{plnbc}. For generality we shall keep $R$ unspecified. For practical purposes, on the other hand, it makes sense to set $R=z_0$, to achieve greatest accuracy.

\subsection{Laplace Transform and Separation of Variables Solution} \label{sec:separation}
The problem's axisymmetry means that in a spherical coordinate formulation there is no dependence on the azimuthal angle, $\theta$. The solution then only depends on $t$, $r$ and $\phi$. With the alternative far-field boundary condition,  \eqref{OuterBC}, the temporal-spatial diffusion problem is amenable to a Laplace transform~/~separation-of-variables solution approach. Defining the Laplace transform of the spatially restricted solutions as,
\begin{equation}
\overline{c}_{\varsigma} (r,\phi:s) = \mathfrak{L}\left\{c_{\varsigma}\right\} = \int_0^{\infty} \e^{-st} c_{\varsigma} (r,\phi,t) \d t, \quad \quad \varsigma = I,II,
\label{lplc}
\end{equation}
the governing equations and boundary conditions become,
\begin{equation}
\left\{
\begin{array}{l}
s \overline{c}_{I} = D_{I}\left[\dfrac{1}{r^2}\dfrac{\dee}{\dee r}\left(r^2\dfrac{\dee \overline{c}_{I}}{\dee r}\right)
+\dfrac{1}{r^2 \sin \phi}\dfrac{\dee}{\dee \phi}\left(\sin \phi \dfrac{\dee \overline{c}_{I}}{\dee \phi}\right) \right]\vspace{0.75cm}\\
s \overline{c}_{II} = D_{II}\left[\dfrac{1}{r^2}\dfrac{\dee}{\dee r}\left(r^2\dfrac{\dee \overline{c}_{II}}{\dee r}\right)
+\dfrac{1}{r^2 \sin \phi}\dfrac{\dee}{\dee \phi}\left(\sin \phi \dfrac{\dee \overline{c}_{II}}{\dee \phi}\right) \right] \vspace{0.75cm}\\
\overline{c}_{I}(R,\phi,s) = \overline{c}_{\infty}(R \cos \phi,s),  \quad \quad \quad r = R, \quad \phi \in[0,\pi], \hsp\\
\left(- D_{II} \dfrac{\dee \overline{c}_{II}}{\dee r}(a-\upsilon,\phi, s)\right) = \alpha \left(- D_{I} \dfrac{\dee \overline{c}_{II}}{\dee r}(a,\phi,s)\right), \quad \quad \phi \in[0,\pi] \hsp \\
\overline{c}_{II}(a-\upsilon,\phi,s) = \beta \overline{c}_{I}(a,\phi,s), \quad \quad \phi \in[0,\pi], \hsp
\end{array}
\right. \label{TrnsfPrblm}
\end{equation}
In  \eqref{TrnsfPrblm}, $s$ is the Laplace parameter complementing the dimensional time $t$, and we emphasize the inclusion of the Laplace transform of  \eqref{OneDsoln},
\begin{equation}
\overline{c}_{\infty}(z,s) = \dfrac{C_0}{s} \e^{\left(-\sqrt{\dfrac{s}{D_I}}(z_0-z)\right)}=\dfrac{C_0}{s} \e^{\left(-\sqrt{z_0^2/D_I}\sqs\right)}\e^{\left(z\sqs/\sqrt{D_I}\right)}, \quad \quad -\infty < z \le z_0. \hsp \label{OneDsolnTr}
\end{equation}
It may readily be confirmed that the finite and unique, separation-of-variables solution of problem  \eqref{TrnsfPrblm} is given by
\begin{multline}
\left\{
\begin{array}{l}
\overline{c}_{I}(r,\phi,s) = \sum_{m=0}^{\infty}\left[A_m i_m\left(\sqrt{\dfrac{r^2}{D_{I}}}\sqs\right)+B_m k_m\left(\sqrt{\dfrac{r^2}{D_{I}}}\sqs\right)\right]P_m(\cos \phi), \quad a \le r \le R,~\phi \in [0,\pi],\\
\overline{c}_{II}(r,\phi,s) = \sum_{m=0}^{\infty}E_m i_m\left(\sqrt{\dfrac{r^2}{D_{II}}}\sqs\right)P_m(\cos \phi), \hspace{2cm} 0 \le r \le a-\upsilon,~\phi \in [0,\pi].\\
\end{array}
\right. \hsp \\ \hsp
\label{TrnsfSln}
\end{multline}
In  \eqref{TrnsfSln} $i_m$ and $k_m$ are the modified spherical Bessel functions of the first and second kind, of order $n$~\citep{abram1965,arfken2001},
\begin{equation}
i_m(z) = \sqrt{\dfrac{\pi}{2z}}I_{m+1/2}(z) \quad ; \quad \quad \quad k_m(z) = \sqrt{\dfrac{2}{\pi z}}K_{m+1/2}(z), \label{sphbssl}
\end{equation}
with $I_{m+1/2}$ and $K_{m+1/2}$ being the fractional order, modified Bessel functions of first and second kind\footnote{Note that some authors~\citep{arfken2001} define $k_m$ differently to that used by others by a multiplicative factor of $\pi/2$.}. Furthermore, $P_m(\cos \theta)$ are the Legendre functions (polynomials) of the first kind \citep{abram1965}.

The expansion coefficients, $A_m$, $B_m$, and $E_m$, giving rise to a unique solution of the Laplace transformed problem are those that satisfy the transformed boundary conditions. To this end, the separation of variables solutions,  \eqref{TrnsfSln}, are substituted into the boundary conditions in  \eqref{TrnsfPrblm}. The resulting equations, involving infinite sums, are reduced to a $3\times3$ linear algebraic system of equations following the application of the orthogonality condition for the Legendre functions. The unique solution to the transformed problem has coefficients, $A_m$ and $B_m$, of the outer solution given by,
\begin{equation}
A_m = \dfrac{sa^2E_m\Lambda_m^{(1)}}{D_{I}} \quad ; \quad B_m = \dfrac{sa^2E_m\Lambda_m^{(2)}}{D_{I}}, \quad \quad \text{for} \quad m \ge 0,
\end{equation}
where
\begin{multline}
\left\{
\begin{array}{l}
\Lambda_m^{(1)} = \dfrac{1}{\beta}
\left[
\dfrac{\beta}{\alpha}\sqrt{\dfrac{D_{II}}{D_{I}}}
k_m\left(\sqrt{a^2/D_{I}}\sqs\right)
i_m^{'}\left(\sqrt{(a-\upsilon)^2/D_{II}}\sqs\right) \right. \\
\hspace{6cm} -  \left.
k_m^{'}\left(\sqrt{a^2/D_{I}}\sqs\right)
i_m\left(\sqrt{(a-\upsilon)^2/D_{II}}\sqs\right)
\right], \\
\Lambda_m^{(2)} = \dfrac{1}{\beta}
\left[
i_m^{'}\left(\sqrt{a^2/D_{I}}\sqs\right)
i_m\left(\sqrt{(a-\upsilon)^2/D_{II}}\sqs\right) \right. \\
\hspace{6cm} - \left.
\dfrac{\beta}{\alpha}\sqrt{\dfrac{D_{II}}{D_{I}}}
i_m\left(\sqrt{a^2/D_{I}}\sqs\right)
i_m^{'}\left(\sqrt{(a-\upsilon)^2/D_{II}}\sqs\right)
\right],\\
\end{array}
\right.\\ \hsp \label{lmda12}
\end{multline}
while the coefficients of the inner solution, $E_m$, are given by
\begin{equation}
\dfrac{a^2s}{D_{I}}E_m = \dfrac{\dfrac{(2m+1)}{2}\Gamma_m}{\left[i_m\left(\sqrt{R^2/D_{I}}\sqs\right)\Lambda_m^{(1)}+k_m\left(\sqrt{R^2/D_{I}}\sqs\right)\Lambda_m^{(2)}\right]}, \quad \text{for} \quad m\ge0,
\label{Emcoef}
\end{equation}
where we have introduced
\begin{eqnarray}
\Gamma_m &=& \int_0^{\pi} \overline{c}_{\infty}(R\cos \phi,t) P_m(\cos \phi) \sin \phi \d \phi, \nonumber \\
&=& \dfrac{2 C_0}{s} \exp{\left(-\sqrt{z_0^2/D_I}\sqs\right)} \int_0^{\pi}\exp{\left(\sqrt{\dfrac{s}{D_I}}R\cos \phi\right)} P_m(\cos \phi) \sin \phi \d \phi. \hsp \label{Gamma_0}
\end{eqnarray}
This last integral may be evaluated directly in a number of ways. However, as a further illustration of the advantage of the reformulated outer boundary condition,  \eqref{OuterBC}, we find it convenient for our purposes to invoke the well-known ``plane-wave'' expansion \citep{arfken2001}
\begin{equation}
\exp{\left(b \cos \varphi\right)} = \sum_{n=0}^{\infty} (2n+1) i_n(b) P_n(\cos \varphi).
\end{equation}
With this expansion in  \eqref{Gamma_0} we find, after another application of the orthogonality condition for the Legendre functions, that
\begin{equation}
\Gamma_m = \dfrac{2C_0}{s}\exp{\left(-\sqrt{z_0^2/D_{I}}\sqs\right)} i_m\left(\sqrt{R^2/D_{I}}\sqs\right). \label{plnwv}
\end{equation}

In order to return to the temporal domain, the solution in the Laplace transformed domain must be inverted. This is achieved through evaluating the complex contour integral,
\begin{equation}
c_{\varsigma}(r,\phi,t) = \mathfrak{L}^{-1}\left\{\overline{c}_{\varsigma}\right\} = \dfrac{1}{2 \pi i} \int_{\sigma - i \infty}^{\sigma + i \infty} \overline{c}_{\varsigma}(r,\phi,s) \e^{st} \d s, \quad \quad \varsigma = I,II. \label{LplcInv_Dfn}
\end{equation}
Of particular interest to us here is the complete, time-dependent solution valid in the cell interior. Inserting our separation-of-variables solution, $\overline{c}_{II}(r,\phi,s)$, into  \eqref{LplcInv_Dfn} we obtain, at least formally,
\begin{equation}
c_{II}(r, \phi, t) = \sum_{m=0}^{\infty} \mathfrak{L}^{-1}\left\{E_m i_m\left(\dfrac{r}{\sqrt{D_{II}}}\sqs\right)\right\} P_m(\cos \phi). \hsp \label{interior}
\end{equation}

\subsection{Small Parameter Expansions} \label{sec:asymptotic}
It is unfortunate that since the Laplace parameter appears in a nontrivial way in the arguments of the modified spherical Bessel functions as well as through multiplicative factors, it is unlikely that an explicit, closed form Laplace inversion is achievable. As an alternative to either a fully numerical solution of the original problem, or a numerical Laplace transform inversion of our separation of variables solution, one may consider limiting interest to either the small-time behaviour of the solution, or its large-time behaviour. These limiting cases correspond to studies of the behaviour of the Laplace transformed solution at large or small values, respectively, of the Laplace parameter $s$. In this paper we focus attention on the large-time behaviour. As a consequence, in this section we shall consider the properties of the transformed solution for values of $s$ in the neighbourhood of $s=0$. This may be readily (but laboriously) achieved. For the solution valid inside the cell, the key result hinges on the $r$-dependent coefficients of the Legendre functions in \eqref{TrnsfSln}, which we here express as infinite series in the parameter $\sqs$. Regrettably, it is necessary to perform a hierarchy of calculations in order to arrive at any useful result. The details of these calculations are summarized in Appendices B and C. One of the two alternative expressions we have derived is
\begin{equation}
E_m i_m\left(\sqrt{r^2/D_{II}}\sqs \right) = C_0\dfrac{D_{I}}{a^2}\dfrac{(2m+1)}{\mathbb{Z}_0^{(m)}}\left(\dfrac{rR}{\sqrt{D_{I}D_{II}}}\right)^{m}s^{(m/2-1)}~\sum_{q=0}^{\infty} \mathbb{S}_q^{(m)}\left(\sqs\right)^q,
\label{TrfAsymSln1}
\end{equation}
where $\left\{\mathbb{Z}_0^{(m)}\right\}$ and $\left\{\mathbb{S}_q^{(m)}\right\}$ are infinite sets of somewhat involved expressions depending on the radius variable, $r$, as well as the system parameters, $a$, $z_0$, $R$, $\upsilon$, $\alpha$, $\beta$, and $\sqrt{D_I/D_{II}}$, but not depending on the Laplace parameter, $s$. This result is improved upon somewhat by the second of our alternative results, namely,
\begin{equation}
E_m i_m\left(\sqrt{r^2/D_{II}}\sqs \right) =  C_0\dfrac{D_{I}}{a^2}\dfrac{(2m+1)}{\mathbb{Z}_0^{(m)}} \left(\dfrac{rR}{\sqrt{D_{I}D_{II}}}\right)^{m} s^{(m/2-1)}~\e^{-\left(\sqrt{z_0^2/D_{I}}\sqs\right)}\sum_{q=0}^{\infty} \mathbb{T}_q^{(m)}\left(\sqs\right)^q,
\label{TrfAsymSln2}
\end{equation}
in which $\left\{\mathbb{T}_q^{(m)}\right\}$ is another, but related, set of $s$-independent coefficients. In this case the full exponential dependence on $z_0$ (from  \eqref{plnwv}) is retained, knowing that an analytic Laplace inversion is possible with this formulation. Details of the derivation of these expansions, and of the ingoing coefficients, are provided in Appendices B and C to this paper.

Adopting these series expansions in  \eqref{interior} we deduce either the convergent series solution,
\begin{eqnarray}
c_{II}(r, \phi, t) = C_0\dfrac{D_{I}}{a^2}\sum_{m=0}^{\infty} \dfrac{(2m+1)}{\mathbb{Z}_0^{(m)}}\left(\dfrac{rR}{\sqrt{D_{I}D_{II}}}\right)^{m} P_m(\cos \phi)~\mathfrak{L}^{-1}\left\{s^{(m/2-1)}\sum_{q=0}^{\infty}\mathbb{S}_q^{(m)}(\sqs)^q\right\}, \hsp \nonumber \\ \hsp\label{TrfAsymSln1_y}
 \end{eqnarray}
or, the alternative series solution,
\begin{eqnarray}
c_{II}(r, \phi, t) &=& C_0\dfrac{D_{I}}{a^2}\sum_{m=0}^{\infty} \dfrac{(2m+1)}{\mathbb{Z}_0^{(m)}}\left(\dfrac{rR}{\sqrt{D_{I}D_{II}}}\right)^{m} P_m(\cos \phi) \nonumber \\
&&\hspace{3.5cm} \times~ \mathfrak{L}^{-1}\left\{s^{(m/2-1)}\e^{-\left(\sqrt{z_0^2/D_{I}}\sqs\right)}\sum_{q=0}^{\infty}\mathbb{T}_q^{(m)}(\sqs)^q\right\}, \hsp \nonumber \\ \hsp \label{TrfAsymSln2_y}
\end{eqnarray}
Based on the difference in the derivations, which appears at the final stage of the calculations, we anticipate that  \eqref{TrfAsymSln1_y} will have a slower rate of convergence at any given time point $t$ than will  \eqref{TrfAsymSln2_y}. For this reason we shall refer to these versions accordingly as the \emph{slower} and \emph{faster convergent series}, respectively. All the same, an important factor in both series is the presence of the combination, $m+q$, of powers of $\sqs$. It should not be surprising that this combination will play an important role, firstly in the form of the inverse Laplace transform, and subsequently in the quality of the asymptotic approximation to the time dependent particle density in the cell interior.

In \ref{sec:largetimesolution} we perform the Laplace inversions indicated in  \eqref{TrfAsymSln1_y} and \eqref{TrfAsymSln2_y} (using standard results summarized in \ref{sec:lapinvser}) to obtain asymptotic time approximations which we then explicitly implement to obtain the solute concentration in the cell interior. The formal expression for the local concentration internal to the cell showing the leading order radial, angular and time dependencies, is thus,
\begin{eqnarray}
c_{II}(r, \phi, t)
&=& C_0\dfrac{D_{I}}{a^2}\dfrac{1}{\sqrt{\pi}}\left\{{ \sqrt{\pi}\dfrac{\mathbb{T}_0^{(0)}}{\mathbb{Z}_0^{(0)}}\left(1-\text{erf}\left(\dfrac{z_0}{\sqrt{4D_It}}\right)\right)} +\dfrac{\e^{\left(-z_0^2/4D_It\right)}}{t^{1/2}}\left\{\left[{ \dfrac{\mathbb{T}_1^{(0)}}{\mathbb{Z}_0^{(0)}}}
+{ 3\left(\dfrac{rR}{\sqrt{D_ID_{II}}}\right)\dfrac{\mathbb{T}_0^{(1)}}{\mathbb{Z}_0^{(1)}}\cos \phi}\right] \right. \right.\nonumber \\
\nonumber \\
&+&\dfrac{z_0}{\sqrt{4D_I}}\dfrac{1}{t}\left[{ \dfrac{\mathbb{T}_2^{(0)}}{\mathbb{Z}_0^{(0)}}}
+ 3\left(\dfrac{rR}{\sqrt{D_ID_{II}}}\right)\dfrac{\mathbb{T}_1^{(1)}}{\mathbb{Z}_0^{(1)}}\cos \phi+ {\dfrac{5}{2}\left(\dfrac{rR}{\sqrt{D_ID_{II}}}\right)^2\dfrac{\mathbb{T}_0^{(2)}}{\mathbb{Z}_0^{(2)}}(3\cos^2 \phi-1)}\right] \nonumber \\
\nonumber \\
&+& \dfrac{1}{2t}\left(\dfrac{z_0^2}{4D_I t}-1\right)\left[{ \dfrac{\mathbb{T}_3^{(0)}}{\mathbb{Z}_0^{(0)}}} +3\left(\dfrac{rR}{\sqrt{D_ID_{II}}}\right)\dfrac{\mathbb{T}_2^{(1)}}{\mathbb{Z}_0^{(1)}}\cos \phi \right. \nonumber \\
\nonumber \\
&+& \left. \left. \left. \dfrac{5}{2}\left(\dfrac{rR}{\sqrt{D_ID_{II}}}\right)^2\dfrac{\mathbb{T}_1^{(2)}}{\mathbb{Z}_0^{(2)}}(3\cos^2 \phi-1) + { \dfrac{7}{2}\left(\dfrac{rR}{\sqrt{D_ID_{II}}}\right)^3\dfrac{\mathbb{T}_0^{(3)}}{\mathbb{Z}_0^{(3)}}(5\cos^3 \phi-3 \cos \phi)}\right] + \cdots \right\} \right\}, \nonumber \\ \nonumber \\ \label{interior_qmGE0_1y}
\end{eqnarray}
where the derived constants $\mathbb{T}_q^{(m)}$ and $\mathbb{Z}_q^{(m)}$ are given explicitly in \ref{sec:seriesexp}.

\section{Solution to the (Corresponding) Spherically Symmetric Diffusion Problem} \label{sec:sphericalcase}
To fully appreciate the analytical results obtained in the foregoing section, it is necessary to draw a comparison with the closely aligned, but greatly simplified, spherically symmetric case of a spherical cell centered within a spherical, exterior region whose boundary at $r=R=z_0$ is a uniform source of nanoparticles, diffusing inwards under a concentration gradient. This is the spherically symmetric system that most closely corresponds to the anisotropic problem described in Section \ref{sec:method}. The governing equations, boundary and initial conditions are, in principle, the same as those presented in Sections \ref{sec:governingeqns} and \ref{sec:bandiconds}, but are greatly simplified as a result of symmetry requirements. In terms of the transformed dependent variable,
\beq
\Theta_{\varsigma} = r  c_{\varsigma}, \quad \varsigma = I,II,
\eeq
the governing equations and boundary conditions (\eqref{diffRegI} - \eqref{plnbc}) reduce to
\begin{equation}
\left\{
\ba{lll}
\dfrac{\partial \Theta_I}{\partial t} = D_I \dfrac{\partial^2 \Theta_I}{\partial r^2}, & \quad & a < r < R ,~t>0\\
\dfrac{\partial \Theta_{II}}{\partial t} = D_{II} \dfrac{\partial^2 \Theta_{II}}{\partial r^2}, & \quad & 0 < r < a- \upsilon,~t>0\\
\Theta_I(R,t) = R C_0, & \quad & t \ge 0,\\
\dfrac{1}{a-\upsilon}\Theta_{II}(a-\upsilon,t) = \dfrac{\beta}{a}\Theta_{I}(a,t), & \quad & t \ge 0,\\
 & & \\
\left.\dfrac{D_{II}}{a-\upsilon}\left(\dfrac{\partial \Theta_{II}}{\partial r}-\dfrac{\Theta_{II}}{r} \right)\right|_{r = a-\upsilon} = \dfrac{\alpha D_I}{a}\left.\left(\dfrac{\partial \Theta_{I}}{\partial r}-\dfrac{\Theta_{I}}{r} \right)\right|_{r=a}, & \quad & t \ge 0,\\
 & & \\
\lim_{r \rightarrow 0}\limits\dfrac{1}{r}\left(\dfrac{\partial \Theta_{II}}{\partial r}-\dfrac{\Theta_{II}}{r} \right) = 0, & \quad & t \ge 0.
\ea
\right.
\end{equation}

As in the anisotropic case, we apply the Laplace transform to these governing equations and conditions. This action leads to an ordinary differential equation system for the transformed dependent variables,
\begin{equation}
\left\{
\ba{lll}
\dfrac{\d^2 \overline{\Theta}_I}{\d r^2} = s D_I \overline{\Theta}_I, & \quad & a < r < R,\\
\dfrac{\d^2 \overline{\Theta}_{II}}{\d r^2} = s D_{II}\overline{\Theta}_{II}, & \quad & 0 < r < a-\upsilon,\\
\dfrac{1}{R}\overline{\Theta}_I(R;s) = \dfrac{C_0}{s}, & \quad & \\
\dfrac{1}{a-\upsilon}\overline{\Theta}_{II}(a-\upsilon;s) = \dfrac{\beta}{a} \overline{\Theta}_{I}(a;s), & \quad & \\
 & & \\
\left.\dfrac{D_{II}}{a-\upsilon}\left(\dfrac{\d \overline{\Theta}_{II}}{\d r}-\dfrac{\overline{\Theta}_{II}}{r} \right)\right|_{r = a-\upsilon} = \dfrac{\alpha D_I}{a}\left.\left(\dfrac{\d \overline{\Theta}_{I}}{\d r}-\dfrac{\overline{\Theta}_{I}}{r} \right)\right|_{r=a}, & \quad & \\
 & & \\
\lim_{r \rightarrow 0}\limits\dfrac{1}{r}\left(\dfrac{\d \overline{\Theta}_{II}}{\d r}-\dfrac{\overline{\Theta}_{II}}{r} \right) = 0,
\ea
\right.\label{sphericalcaseLT}
\end{equation}
where
\begin{equation}
\overline{\Theta}_{\varsigma}(r;s) = \int_0^{\infty}\e^{-st}~\Theta_{\varsigma}(r,t)~ \d t, \quad \varsigma = I,II,
\end{equation}
with $s$ again as the complementary Laplace parameter (to time t).

Our primary interest lies with the cell interior region (the storage vacuole). The solution to (\ref{sphericalcaseLT}.ii) satisfying the boundary conditions Eqs (\ref{sphericalcaseLT}.iii-vii) is
\begin{equation}
\overline{\Theta}_{II}(r;s) = \dfrac{\beta C_0 D_{I}^{1/2} R(1-\upsilon/a)}{\Psi(p)}\,\dfrac{\sinh{\left(s^{1/2}r{\big/}D_{II}^{1/2}\right)}}{s^{1/2}} \quad \left( = r \overline{c}_{II} \right), \label{sphericalcaseSLN}
\end{equation}
with
\begin{eqnarray}
\Psi(s) &=& \left(1-\dfrac{D_{II}}{D_I}\dfrac{\beta}{\alpha}\dfrac{1}{(1-\upsilon/a)}\right)\sinh\left({s^{1/2}a}{\big/}{D_{II}^{1/2}}\right)\sinh\left(s^{1/2}(R-a){\big/}{D_{I}^{1/2}}\right) \qquad \nonumber \\
&& \qquad + \quad \dfrac{s^{1/2}a}{D_I^{1/2}} \Bigl[\left(\dfrac{D_{II}}{D_I}\right)^{1/2}\dfrac{\beta}{\alpha}\cosh\left({s^{1/2}a}{\big/}{D_{II}^{1/2}}\right) \sinh\left(s^{1/2}(R-a){\big/}{D_{I}^{1/2}}\right) \nonumber\\
&& \qquad \qquad \qquad + \quad \left. \sinh\left({s^{1/2}a}{\big/}{D_{II}^{1/2}}\right) \cosh\left(s^{1/2}(R-a){\big/}{D_{I}^{1/2}}\right) \right]. \label{sphericalcaseROOT}
\end{eqnarray}
From the inverse Laplace transform, the time dependent solution in the cell interior is then,
\begin{equation}
\Theta_{II}(r,t) = \dfrac{1}{2\pi i}\int_{\gamma-i\infty}^{\gamma+i\infty} \e^{st}~\overline{\Theta}_{II}(r;s)~ \d s.
\end{equation}
We use residue theory to evaluate this complex integral. In this case we look for the zeros of the denominator of \eqref{sphericalcaseSLN}. Despite the pervasive presence of the factor $s^{1/2}$ throughout \eqref{sphericalcaseSLN} and \eqref{sphericalcaseROOT}, $s=0$ is not a branch point as a result of the pairwise multiplicative appearances of $s^{1/2}$ in every term, which effectively leads to all the expressions possessing only even powers of $s^{1/2}$. For this reason, a small-$s$ calculation to derive an asymptotic time dependent concentration, analogous to that pursued in the anisotropic problem, proves fruitless; the finite end result is only the steady state solution $c_{II}=\beta C_0$ (which in and of itself already suggests an exponential time dependence rather than an algebraic time dependence.) The zeros of the denominator are in fact simple poles which, for integral convergence, appear on the negative real $s$ axis. For this reason it proves useful to change to a new parameter: $s = - q^2$.

We look for zeros of Eqs \eqref{sphericalcaseSLN} and \eqref{sphericalcaseROOT}. We readily identify $s=q=0$ as one root, which contributes with the steady-state solution. The remaining, non-trivial roots are the zeros of
\begin{equation}
\begin{array}{r}
\Phi(q) = \Psi\left(s=-q^2\right) = -\left(1-\dfrac{D_{II}}{D_I}\dfrac{\beta}{\alpha}\dfrac{1}{(1-\upsilon/a)}\right)\sin\left({qa}{\big/}{D_{II}^{1/2}}\right)\sin\left(q(R-a){\big/}{D_{I}^{1/2}}\right) \\
- \dfrac{qa}{D_I^{1/2}} \Bigl[\left(\dfrac{D_{II}}{D_I}\right)^{1/2}\dfrac{\beta}{\alpha}\cos\left({qa}{\big/}{D_{II}^{1/2}}\right) \sin\left(q(R-a){\big/}{D_{I}^{1/2}}\right) \qquad \qquad\\
+  \sin\left({qa}{\big/}{D_{II}^{1/2}}\right) \cos\left(q(R-a){\big/}{D_{I}^{1/2}}\right) \Bigr]=0.
\end{array} \label{sphericalcaseQROOT}
\end{equation}
This equation produces two classes of solution. Note first that although there is a set of $q$-values for which $\sin({qa}{\big/}{D_{II}^{1/2}})=0$ and there is a set for which $\sin(q(R-a){\big/}{D_{I}^{1/2}})=0$, members of neither set are necessarily zeros of $\Phi(q)$. However, should the physical conditions be such that the periods of these sine functions are commensurate, \textit{i.e.}, should values of $a$, $R$ and $D_{II}/D_I$ be such that integers $M$ and $L$ can be found to satisfy
\begin{equation}
\dfrac{M}{\left(D_{II}/D_I\right)^{1/2}} = L\dfrac{R-a}{a}, \label{CommensuratePeriods}
\end{equation}
then there \emph{will} be a common \emph{subset} of the above-mentioned sets for which both sine functions vanish. This set comprises discrete roots $q_n^{(1)}$ such that
\begin{equation}
\dfrac{a}{\sqrt{D_{II}}}\dfrac{q_n^{(1)}}{L} = n \pi = \dfrac{(R-a)}{\sqrt{D_I}}\dfrac{q_n^{(1)}}{M}, \qquad n = 1,2,3, \ldots. \label{sphericalcaseQROOT1}
\end{equation}
which are thus zeros of $\Phi(q)=0$. Independent of this class of roots (which does not exist under incommensurate physical conditions) are the roots of
\begin{equation}
\left(1-\dfrac{D_{II}}{D_I}\dfrac{\beta}{\alpha}\dfrac{1}{(1-\upsilon/a)}\right) + \dfrac{qa}{D_I^{1/2}}~\left[ \left(\dfrac{D_{II}}{D_I}\right)^{1/2}\dfrac{\beta}{\alpha}\cot\left({qa}{\big/}{D_{II}^{1/2}}\right)+\cot\left(q(R-a){\big/}{D_{I}^{1/2}}\right) \right]=0.
\end{equation}
As both $\cot(q(R-a){\big/}{D_{I}^{1/2}})$ and $\cot({qa}{\big/}{D_{II}^{1/2}})$ are periodic functions, now with incommensurate periods, the intersection of their scaled sum with $-\dfrac{1}{q}\left(1-\dfrac{D_{II}}{D_I}\dfrac{\beta}{\alpha}\dfrac{1}{(1-\upsilon/a)}\right)$ occurs at a countably infinite number of discrete points. In the absence of closed form expressions for these roots, we identify members of this discrete infinite set simply as $q_m^{(2)},~m=1,2,3,\ldots$.

With the (at most) two sets of discrete roots of $\Phi(q)$, which are simple poles of $\overline{\Theta}_{II}$, the time-dependent, dimensionless concentration is given by the inverse integral transform,
\begin{eqnarray}
\Theta_{II}(r,t) &=& \dfrac{1}{2\pi i}\int_{\gamma - i \infty}^{\gamma + i \infty} \e^{st}~\overline{\Theta}_{II}(r;s)~\d s = \sum_{k=1}^{\infty} Res\left(\e^{st}\overline{\Theta}_{II},s_k\right), \nonumber \\
&=& \beta C_0 r - \sum_{m=1}^{\infty} \dfrac{\beta R C_0(1-\upsilon/a)\sin(q_m^{(2)}r/D_{II}^{1/2})\sin(q_m^{(2)}a/D_{II}^{1/2})\sin(q_m^{(2)}(R-a)/D_I^{1/2})}{\psi(q_m^{(2)})}\e^{-(q_m^{(2)})^2 t} \nonumber \\
&& \qquad \qquad \qquad + \sum_{n=1}^{\infty} \dfrac{2\beta R C_0(1-\upsilon/a)\sin(n\pi L r/a)(-1)^{n(M+L)}}{n\pi(\left(D_{II}/D_I\right)^{1/2}M+L)}\e^{-n^2\pi^2L^2D_{II}t/a^2}, \label{sphericalcaseFINALSOLN}
\end{eqnarray}
where
\begin{eqnarray}
\psi(q_m^{(2)}) &=& q_m^{(2)}\left((R/a-1)\sin^2(q_m^{(2)}a/D_{II}^{1/2})+ \dfrac{\beta}{\alpha}\sin^2(q_m^{(2)}(R-a)/D_I^{1/2})\right) \nonumber \\
&& \qquad \qquad + \dfrac{1}{q_m^{(2)}}\left(1-\dfrac{D_{II}}{D_I}\dfrac{\beta}{\alpha}\dfrac{1}{1-\upsilon/a}\right)\sin^2(q_m^{(2)}a/D_{II}^{1/2})\sin^2(q_m^{(2)}(R-a)/D_I^{1/2}),
\end{eqnarray}
and where the second series in \eqref{sphericalcaseFINALSOLN} is \emph{not} present if the physical conditions do not result in commensurate periods, \eqref{CommensuratePeriods}. The time-independent term in \eqref{sphericalcaseFINALSOLN} is the steady state, uniform concentration solution.

\section{Cellular Uptake of Diffusive Particles} \label{sec:discussion1}
Our principal interest is with the time-dependent accumulation of nanoparticles in the interior of a spherical cell due to an influx from an external environment. This was similarly of interest to previous authors~\citep{crank1956,mild1971}. Here we analyze the particle uptake assuming a unidirectional source and, for comparison, a spheroidal source. The relevant quantity for this study is the time-dependent total particle amount, $\mathfrak{M}(t)$. This is arguably a more meaningful, experimentally-relevant quantity. That is, it is more likely to be a quantity that can be measured, than is the time dependent local concentration, $c_{II}(r,\phi,t)$.

\subsection{The Rate of Nanoparticle Uptake}
Under the physical model conditions set for both the anisotropic problem and the spherically symmetric problem, it has been shown (see also \ref{sec:stdystte}) that at steady state the interior of the cell will be filled to the uniform concentration of $\beta C_0$, in which case the total number of particles will then be
\begin{equation}
\mathfrak{M}_{\infty} = \dfrac{4}{3}\pi (a-\upsilon)^3 \beta C_0. \label{inftot}
\end{equation}
Letting $\mathfrak{M}(t)$ denote the total number of particles in the cell at any finite point in time, we may readily deduce that the rate of change of $\mathfrak{M}(t)$ is given by the vector surface integral of the local particle flux entering the cell~\citep{miklavcic2020},
\begin{equation}
\dfrac{\d \mathfrak{M}}{\d t}(t) = \oiint_{cell} \left(-D_{II}\del c_{II}(\mb{r})\right) \cdot \d \mb{S}, \label{ratetot1}
\end{equation}
and the total particle number at any point in time, $t$, is then
\begin{equation}
\mathfrak{M}(t) = \int_0^{t} \dfrac{\d \mathfrak{M}}{\d t'}(t') ~\d t'. \label{tautot1}
\end{equation}
\subsection{Particle Uptake from a Unidirectional Source} \label{sec:particleuptake1D}
As the cell is spherical, for the anisotropic case \eqref{ratetot1} simplifies somewhat to leave an integral over the azimuthal angle,
\begin{equation}
\dfrac{\d \mathfrak{M}_U}{\d t'}(t') = 2 \pi D_{II} (a-\upsilon)^2 \int_{0}^{\pi} \dfrac{\partial c_{II}}{\partial r} (a-\upsilon,\phi,t') \sin \phi ~\d \phi. \label{ratetot2}
\end{equation}
Given that we cannot perform the inverse Laplace transform to determine $c_{II}(r,\phi,t)$, we cannot complete the above calculation to give a result that will be valid for all times. However, we do have the asymptotic expression for the local concentration inside the cell. Consequently, it \emph{is} feasible to study the \emph{approach} to steady state. Rewriting  \eqref{tautot1} we deduce a suitable quantity:
\begin{eqnarray}
\mathfrak{M}_U(t) &=& \left(\int_0^{\infty} - \int_{t}^{\infty} \right)\dfrac{\d \mathfrak{M}_U}{\d t'}(t') \d t' \hsp \nonumber \\
&=&  \mathfrak{M}_{\infty} - 2 \pi D_{II} (a-\upsilon)^2 \int_{t}^{\infty} \int_{0}^{\pi} \dfrac{\partial c_{II}}{\partial r} (a-\upsilon,\phi,t') \sin \phi ~\d \phi ~\d t'. \label{tautot2}
\end{eqnarray}
With the normalized result,
\begin{equation}
\dfrac{\mathfrak{M}_U(t)}{\mathfrak{M}_{\infty}} = 1 - \dfrac{3 D_{II}}{2 \beta C_0 (a-\upsilon)} \int_{t}^{\infty} \int_{0}^{\pi} \dfrac{\partial c_{II}}{\partial r} (a-\upsilon,\phi,t') \sin \phi ~\d \phi ~\d t', \label{tautot3}
\end{equation}
we have a measure of the approach to steady state content (the time integral approaches zero as $t \rightarrow \infty$). For $t$ sufficiently large the format of  \eqref{tautot3} permits the use of our asymptotic solution in the double integral.

It is useful to make the following observations. Irrespective of the level of approximation assumed any term in $c_{II}$ that is independent of radius, $r$, will \emph{not} contribute to the total. Moreover, \emph{all} terms in $c_{II}$ involving a Legendre polynomial, $P_m(\cos \phi)$, with the exception of the case $m=0$, will result in a zero angle integral, by the orthogonality property of the Legendre functions (all terms apart from the first are orthogonal to the constant, $m=0$ term). Consequently, the only contributions to the time-dependent, total particle number will be the subset of spatially dependent, $m=0$ terms, which were singled out for special interest in \ref{sec:meq0qGE0}. Of these the term $m=q=0$ is spatially independent, as shown in \ref{sec:stdystte}. Therefore, the only contributing terms will be those in  \eqref{interior_meq0qGE0_2x} for which $m=0$ and $q \ge 1$.\footnote{One possible physical interpretation of the limited contributions, is that the angle-dependent terms represent a balance between the diffusion of particles into and out of the cell, resulting in a net zero change in total number.}

Focusing thus on our key result, we insert our approximation \eqref{interior_qmGE0_1y} in  \eqref{tautot3} and perform the necessary operations. We get (noting in the process the cancelation of the following factors: $\beta C_0$, $(a-\upsilon)$, $D_{II}$, and 2 and 3),
\begin{eqnarray}
\dfrac{\mathfrak{M}_U(t)}{\mathfrak{M}_{\infty}} &=& 1 -\text{erf}\left(\dfrac{z_0}{\sqrt{4D_It}}\right) + \dfrac{(4D_I)}{z_0^2}\left\{\dfrac{1}{6}\dfrac{R^2}{D_I}\left[1+\dfrac{3}{5}\dfrac{a^2}{R^2}\dfrac{D_I}{D_{II}}(1-\epsilon)^2\right] - \dfrac{\mathbb{Z}_2^{(0)}}{\mathbb{Z}_0^{(0)}}\right\} \hsp \nonumber \\
&\times& \left\{\dfrac{1}{8}\text{erf}\left(\dfrac{z_0}{\sqrt{4D_It}}\right) - \dfrac{1}{4 \sqrt{\pi}}\left(\dfrac{z_0}{\sqrt{4D_It}}\right)\e^{\left(-\dfrac{z_0^2}{4D_It}\right)}\left[1-2\left(\dfrac{z_0^2}{4D_It}\right)\right]\right\}+\ldots,
\hsp \label{MonMinf}
\end{eqnarray}
where $\mathbb{Z}_2^{(0)}/\mathbb{Z}_2^{(0)}$ is given in  \eqref{Z20onZ00}.

Of the factors contributing to $\mathfrak{M}_U(t)/\mathfrak{M}_{\infty}$, the scaling of the third term indicates that the approach to the expected total is delayed the further away is the source from the cell, and the slower is the diffusion in the external medium. The same can be said of their effects through the error function and the exponential function. The constants $\alpha$ and $\beta$ play less prominent roles in the rate of approach to steady state, a feature shared with the internal diffusion constant $D_{II}$.

Expressed explicitly in \emph{dimensionless} system properties and parameters (for numerical purposes),
\begin{equation}
u_{\varsigma} = \dfrac{c_{\varsigma}}{C_0}, \text{ \ \ \ }\varsigma = I,II \quad ; \quad \rho = \dfrac{r}{a} \quad ; \quad X = \dfrac{R}{a}=\dfrac{z_0}{a} \quad ; \quad \epsilon= \dfrac{\upsilon}{a} \quad ; \quad \tau = \dfrac{D_I t}{a^2}, \label{nondim_param}
\end{equation}
we have from \eqref{MonMinf} the main objective of this paper: the relative difference,
\begin{eqnarray}
\dfrac{\mathfrak{M}_{\infty}-\mathfrak{M}_U(\tau)}{\mathfrak{M}_{\infty}} &=& \text{erf}\left(\dfrac{X}{\sqrt{4\tau}}\right) - \dfrac{1}{6}\dfrac{1}{X^2}\left\{\dfrac{2}{5}\dfrac{1}{A}(1-\epsilon)^2+\left[\left(\dfrac{\beta}{\alpha}(1-\epsilon)-1\right)\left(1-2\dfrac{1}{X}\right) - 2\right]\right\} \hsp \nonumber \\
&\times& \left\{\dfrac{1}{2}\text{erf}\left(\dfrac{X}{\sqrt{4\tau}}\right) - \dfrac{1}{ \sqrt{\pi}}\left(\dfrac{X}{\sqrt{4\tau}}\right)\e^{\left(-\dfrac{X^2}{4\tau}\right)}\left[1-2\left(\dfrac{X^2}{4\tau}\right)\right]\right\}+\ldots,
\hsp \label{MonMinf2}
\end{eqnarray}
where we have assumed $R=z_0$ only in the term within the first pair of square brackets.

\subsection{Particle Uptake from a Spherically Symmetric Source}
To compare with the anisotropic situation we now turn attention to the rate at which nanoparticles accumulate from a spherically symmetric nanoparticle source a distance $R$ from the cell center, $\mathfrak{M}_S(\tau)$. Corresponding to \eqref{MonMinf2} and using \eqref{sphericalcaseFINALSOLN} with dimensionless variables and parameters, Eq \eqref{nondim_param}, we have
\begin{eqnarray}
\dfrac{\mathfrak{M}_{\infty}-\mathfrak{M}_S(\tau)}{\mathfrak{M}_{\infty}} &=& \dfrac{3D_{II}}{\beta C_0(a-\upsilon)}~\int_{t}^{\infty} \left.\dfrac{\partial c_{II}}{\partial r}\right|_{r=a-\upsilon} \d t'= 3A~\int_{\tau}^{\infty} \left.\dfrac{\partial}{\partial \rho}\left(\dfrac{1}{\beta(1-\epsilon)}~\dfrac{\Theta_{II}}{\rho} \right) \right|_{\rho=1-\epsilon} \d \tau', \nonumber \\
\nonumber \\
&=& \sum_{n=1}^{\infty} \dfrac{6AX(-1)^{n(M+L)}}{n^3\pi^3L^2A(A^{1/2}M+L)} \left[\dfrac{n\pi L\cos(n\pi L(1-\epsilon))}{(1-\epsilon)}-\dfrac{\sin(n\pi L(1-\epsilon))}{(1-\epsilon)^2}\right]\e^{-n^2\pi^2L^2A \tau}. \nonumber \\
&-&~\sum_{m=1}^{\infty} \dfrac{3AX\sin(q_m^{(2)}/A^{1/2})\sin(q_m^{(2)}(X-1))}{\psi(q_m^{(2)})(q_m^{(2)})^2} \nonumber \\
\nonumber \\
&& \qquad \qquad \qquad \qquad \times \left[\dfrac{q_m^{(2)}\cos(q_m^{(2)}(1-\epsilon)/A^{1/2})}{A^{1/2}(1-\epsilon)}-\dfrac{\sin(q_m^{(2)}(1-\epsilon)/A^{1/2})}{(1-\epsilon)^2}\right]\e^{-(q_m^{(2)})^2 \tau} \nonumber \\
\nonumber \\
\label{MonMinf2_sphericalcase_all}
\end{eqnarray}

Keeping only the first (dominating) exponential term of each series in \eqref{MonMinf2_sphericalcase_all} (assuming commensurate conditions apply), \eqref{MonMinf2} may be contrasted with
\begin{eqnarray}
\dfrac{\mathfrak{M}_{\infty}-\mathfrak{M}_S(\tau)}{\mathfrak{M}_{\infty}}
&=& \dfrac{6AX(-1)^{(M+L)}}{\pi^3L^2A(A^{1/2}M+L)} \left[\dfrac{\pi L\cos(\pi L(1-\epsilon))}{(1-\epsilon)}-\dfrac{\sin(\pi L(1-\epsilon))}{(1-\epsilon)^2}\right]\e^{-\pi^2L^2A \tau}. \nonumber \\
&-&~\dfrac{3AX\sin(q_1^{(2)}/A^{1/2})\sin(q_1^{(2)}(X-1))}{\psi(q_1^{(2)})(q_1^{(2)})^2} \nonumber \\
\nonumber \\
&& \qquad \qquad \qquad \qquad \times \left[\dfrac{q_1^{(2)}\cos(q_1^{(2)}(1-\epsilon)/A^{1/2})}{A^{1/2}(1-\epsilon)}-\dfrac{\sin(q_1^{(2)}(1-\epsilon)/A^{1/2})}{(1-\epsilon)^2}\right]\e^{-(q_1^{(2)})^2 \tau} \nonumber \\
\nonumber \\
\label{MonMinf2_sphericalcase}
\end{eqnarray}

\begin{figure}[t!]
\vspace{-1.0cm}
\includegraphics[width=0.9\columnwidth]{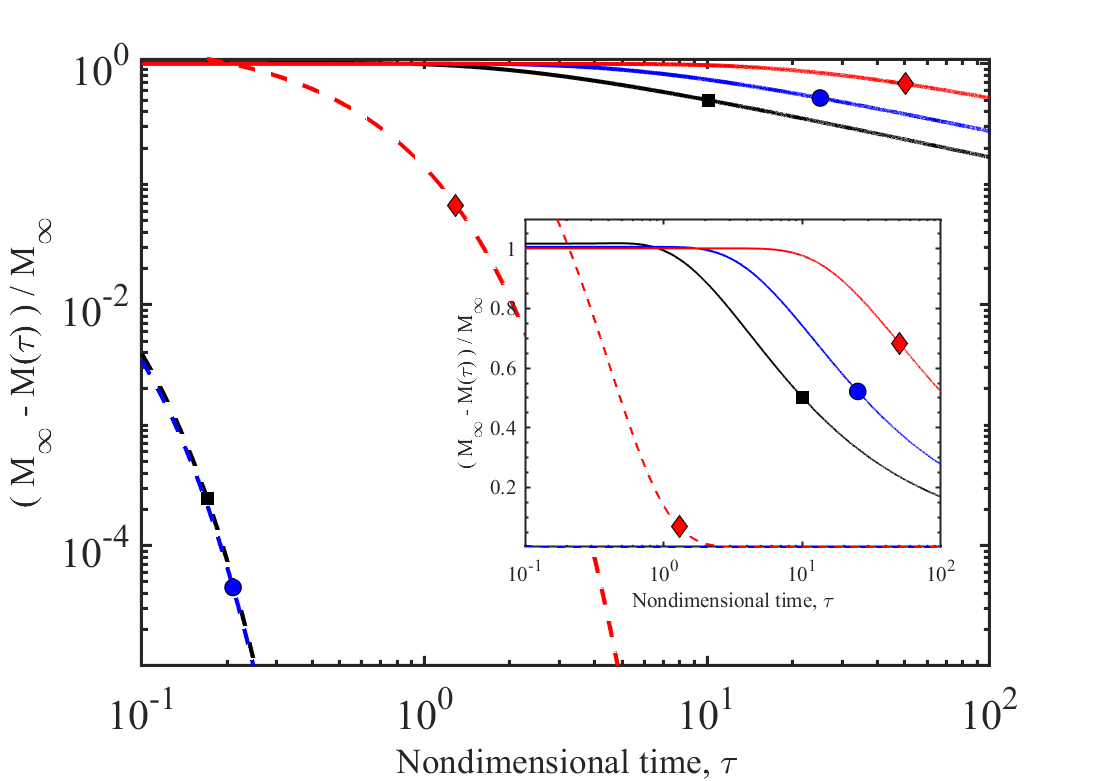}
\caption{Comparison of time asymptotic approximations to the rate of accumulation of nanoparticles by a spherical cell under two different exposure conditions. Solid lines refer to the asymptotic approach to steady state under anisotropic conditions expressed by Eq \eqref{MonMinf2}, while dashed lines refer to the sum of the leading order exponential terms given in Eq \eqref{MonMinf2_sphericalcase} under spherically symmetric exposure. The symbol identifiers refer to different distances between the cell boundary and the source: $X=3$ - black squares; $X=5$ - blue circles; and $X=10$ - red diamonds. Other parameter conditions: $A=4$, $\alpha=\beta=1$, $\delta=0$. The parameter choices satisfy the commensurate requirement in the spherically symmetric case with $L=1$ and $M=4,8,~\&~18$, respectively, for the three $X$ values. In order to show all six time profiles in the one figure, the results are plotted in the main figure on a log-log scale. The inset shows the same data (minus the spherically symmetric results for $X=3~\&~5$) on a log-linear scale. \label{fig:figure2}}
\end{figure}

\subsection{Comparison of Particle Uptake Rates}
Quantities \eqref{MonMinf2} and \eqref{MonMinf2_sphericalcase} measure the long time approach to the steady state condition of a fully laden cell. That is, these quantities measure the respective rates at which a cell accumulates nanoparticles under two different exposure conditions. While one might assume that the rate at which nanoparticles are taken up by a cell is primarily dependent on its intrinsic properties, even a casual examination of the above two results reveals that instead the accumulation rate is primarily dependent on the mode of particle delivery to the cell.

To leading order \eqref{MonMinf2} (as well as \eqref{MonMinf}) expresses the fact that nanoparticle accumulation is foremost dictated by the rate (and manner!) by which particles arrive at the cell's location. This is then modulated through a scaling of the time dependencies by coefficients possessing information about the cell: its size ($a$), internal diffusion constant ($D_{II}$), membrane thickness ($\epsilon$), membrane flux resistance ($\alpha$) and the cell's overall propensity to accommodate particles ($\beta$).  The latter characteristics do not alter the time dependencies per se. What is particularly clear from \eqref{MonMinf2} is the explicit role played by the unidirectional character of nanoparticle exposure through the erf function (which features in the solution of the unidirectional diffusion problem).

To further illustrate the difference between the two model results we present in Figure \ref{fig:figure2} a numerical comparison for the idealized case of $\alpha=\beta=1$ and $\epsilon=0$ (the case of boundary continuity of concentration and flux and of an infinitesimally thin cell membrane). We assume the conditions of $A=4$ (the diffusion rate interior to the cell is 4 times as fast as that exterior to the cell) and $X=3,5,~\&~10$ (the shortest distance between the cell surface and the source of particles is $X-1=2,4,~\&~9$ multiples of cell radius. Under these conditions Eq \eqref{CommensuratePeriods} is satisfied with $L=1$ and $M=4,8,~\&~18$, respectively. In the main figure the results are presented on a log-log plot in order to show all six cases in the one figure. The inset shows the same data plotted on log-linear scale. If not already appreciated from the contrasting functional dependencies displayed by \eqref{MonMinf2} and \eqref{MonMinf2_sphericalcase}, the numerical comparison in Figure \ref{fig:figure2} drives home the extent to which the cases differ in their predictive rates. For the same cell properties, a uniform exposure from a spherically symmetric source of nanoparticles results in a much faster (exponentially fast) accumulation of particles compared to the much slower particle accretion from exposure from one direction.

The reader is reminded that the expressions plotted are leading order approximations for the long time approach to steady state. Consequently, the numerical results should not expected to be accurate at short times, and no meaning should be assigned to said results.

\section{Concluding Remarks} \label{sec:conclusion}
In this paper we addressed the problem of a unidirectional diffusive transport of particles toward, around and through a spherical cell. This problem has not been studied previously in an analytic form to any rigorous extent. Most diffusion problems are amenable to analytical solution when confined to a simple and consistent geometries. Examples of the latter are the works of Crank \citep{crank1956}, Philip \citep{philip1964} and Mild \citep{mild1971}, which have inspired the present calculation. The difficulty in reconciling different geometric conditions (in the present case that of one Cartesian dimension and one spherical geometry) which has hindered previous attempts, is here overcome partly through an implementation of an effective, so-called plane wave boundary condition, and partly through an asymptotic, large-time analysis. In the latter respect our efforts mirror, but simultaneously generalize, the asymptotic calculations of \citep{philip1964} and \citep{mild1971}.

It is worthwhile repeating that although a numerical solution of the original partial differential equation may be readily obtained (say, using a standard time-dependent, finite element approach) there is limited analytical understanding of the dependence on the model parameters to be gained with such an approach, beyond that which may be deduced from purely physical arguments. On the other hand, symbolic computation tools may help in an explicit asymptotic analysis (as we ourselves have gleaned) but these are limited in their ability to generate arbitrary orders of terms of the asymptotic approximation. At the very least, considerable user input is needed to develop a useful solution expansion. Although the calculations presented here are reasonably straightforward but tedious, the results are general and can be exploited to arbitrary order, at least in theory. As already mentioned, the hierarchy of factors can be coded easily into a numerical algorithm to study diffusion behaviour in different physical cases. This is one goal of a separate publication wherein we compare the asymptotic results with a full numerical solution of the diffusion problem by finite element method~\cite{Kumar2024}.

The rate of particle accumulation in a cell is a quantity that is possible to monitor experimentally, at least qualitatively~\cite{Limbach2005,Cui2016} and as such is of greater experimental value to model than is the local, time-dependent particle concentration. Deriving the time-dependent accumulation of nanoparticles was the main goal of this paper. As demonstrated both analytically and numerically, for the one same cell, the rate is found to be greatly dependent on the mode of exposure. Arguably, it is exposure from one direction that is the more physically relevant and the more likely encountered situation, rather than the more artificial scenario of a spherically symmetric exposure. Our result can thus be considered a more applicable contribution, despite being an asymptotic approximation. This is demonstrated in a comparison of theory with experimental results in \cite{Miklavcic2024b}.

Equation \eqref{MonMinf2} is based on the assumption of an isolated cell. It is thus reasonable to expect that its relevance would also extend to a low solution concentration of cells. If the cells are uniformly distributed in the suspension, a direct statistical analysis could account for the different distances ($z_0$) to a nanoparticle source. An attractive prospect for future study is the case of a nonuniform cell membrane response to nanoparticles. The simplest way in which this may be achieved is by adopting prescribed, static and angle-dependent boundary conditions. A more ambitious possibility is that of a non-uniform membrane flux that is dependent on local concentrations and thus adapts with time around the membrane~\citep{foster2019}. An active role of the membrane thickness, $v$, has not been examined here. Its potential influence may also be explored in a future study by modeling that region as a chemically reactive matrix~\citep{AlObaidi2015,miklavcic2014}. For charged nanoparticles (not assumed here), one may consider changes in the permeability of the membrane via a electroporation mechanism based on the difference in membrane potentials on each side of the cell membrane~\citep{Weaver2000,Leguebe2014}. However, the more immediate consideration for a future work is to extend the present model to include the effect of nanoparticle sedimentation, which has been shown in \cite{Limbach2005} and \cite{Cui2016} to be significant for larger nanoparticles.

\section*{Conflict of Interest Statement}
The author declares that the research was conducted in the absence of any commercial or financial relationships that could be construed as a potential conflict of interest.

\section*{Author Contributions}
The author was solely responsible for the model development, mathematical methodology, analysis and drafting of the paper.\par

\section*{Funding}
This project is supported by the Australian Research Council (Discovery project grant DP200103168).\par

\section*{Acknowledgments}
The author would like to thank Dr Jorge Aarao and Mr Sandeep Santosh Kumar for many useful discussions.\par

\bibliography{Miklavcic_I_2024_arXiv}

\newpage

\appendix
\renewcommand{\theequation}{A.\arabic{equation}}
\section*{Appendices}
\section{Addressing Sedimentation}
Suppose the general problem concerning the mass transport of particles occurs by diffusion and sedimentation which is described generally by the equation
\begin{equation}
\dfrac{\partial c}{\partial t} = D \nabla_x^2 c - \mathbf{v}\cdot \nabla_x c, \label{general}
\end{equation}
a version of which would be applicable in the external medium and a version applicable in the cell interior. The dependent variable $c(\mathbf{r},t)$ is the space and time-dependent particle mass concentration, $D$ is the Stokes-Einstein diffusion constant for the medium, and $\mathbf{v}$ is the particle sedimentation velocity (which might, say, be directed in the negative $z$-direction: $\mathbf{v}=-v\mathbf{k}$). The constant sedimentation speed $v$ is established by the force balance of buoyancy and Stokes's drag:
\begin{equation*}
v = \dfrac{2g(\rho_p-\rho_m)r_p^2}{9 \eta},
\end{equation*}
where $g$ is the gravitational acceleration, $\rho_{p(m)}$ is the mass density of particles(medium), $\eta$ is the viscosity of the medium and $r_p$ is the hydrodynamic radius of the (spherical) particles.

For a constant sedimentation velocity, consider the independent variable transformation,
\begin{equation}
\xi_i = v_i t - x_i, \quad i=1,2,3; \quad \tau = t \label{trass}
\end{equation}
with $c(\mathbf{r},t) = \overline{c}(\mathbf{\xi},\tau)$, from which we can derive
\begin{eqnarray}
\dfrac{\partial c}{\partial t} &=& \dfrac{\partial \overline{c}}{\partial \tau} + \sum_j v_j \dfrac{\partial \overline{c}}{\partial \xi_j},\\
\dfrac{\partial c}{\partial x_i} &=& - \dfrac{\partial \overline{c}}{\partial \xi_i}.
\end{eqnarray}
Consequently,  \eqref{general} transforms as
\begin{equation}
\dfrac{\partial c}{\partial t} - D \nabla_x^2 c + \mathbf{v}\cdot \nabla_x c = \dfrac{\partial \overline{c}}{\partial \tau} - D \nabla_{\xi}^2 \overline{c}
 = 0.
\end{equation}
Thus, the problem \emph{with} sedimentation may be solved using the solution of the problem \emph{without} sedimentation upon invoking the inverse transformation to  \eqref{trass} (with suitable consideration of altered boundary conditions).

\renewcommand{\theequation}{B.\arabic{equation}}
\section{Bessel Function Expansions}
From the outset we aim to present the separation of variables solution in a form that is amenable to Laplace inversion. To this end we develop below series expansions of each contributing function of the Laplace parameter, $s$, and subsequently combine these into a single series expression in, specifically, $\sqs$, which we may invert using standard complex integration techniques.

\subsection{Generic Results}
We begin by stating the series representations of the modified spherical Bessel functions $i_m$ and $k_m$ and their derivatives $i_m^{'}$ and $k_m^{'}$~\citep{abram1965,arfken2001}.\footnote{Some care should be exercised due to a difference in definition of $k_m$~\citep{arfken2001}, involving a factor of $\pi/2$.} The ascending series for $i_m$ and $i_m^{'}$ are
\begin{multline}
\left\{
\begin{array}{l}
i_m(z) = z^m\sum_{j=0}^{\infty}\dfrac{\left(z^2/2\right)^j}{j!(2m+2j+1)!!} = z^m\sum_{j=0}^{\infty}\dfrac{z^{2j}}{j!(2m+2j+1)!! 2^j } \doteq \sum_{j=0}^{\infty} d_j^{(m)} z^{2j+m}. \\
\\
i_m^{'}(z) = \sum_{j=0}^{\infty} (2j+m) d_j^{(m)} z^{2j+m-1}.
\end{array}
\right. \hsp \label{ibsslpr}
\end{multline}
The last equality in the first of  \eqref{ibsslpr} defines the coefficient $d_j^{(m)}$ to appear in later equations. It is important to note that these two series are absolutely and uniformly convergent. Although divergent at zero arguments, ascending-like series for $k_m$ and $k_m^{'}$ can also be developed:
\begin{eqnarray}
k_m(z) &=& \dfrac{\e^{-z}}{z} \sum_{j=0}^{m}\dfrac{(m+j)!}{j!(m-j)!}\dfrac{1}{(2z)^j} = \dfrac{\e^{-z}}{z^{m+1}} \sum_{j=0}^{m}\dfrac{(2m-j)!}{j!(m-j)!}\dfrac{z^j}{2^{m-j}}, \hsp \nonumber \\
&=& \dfrac{1}{z^{m+1}}\left(\sum_{l=0}^{\infty}\dfrac{(-1)^lz^l}{l!}\right) \left(\sum_{j=0}^{m}\dfrac{(2m-j)!}{j!(m-j)!}\dfrac{z^j}{2^{m-j}}\right) = \dfrac{1}{z^{m+1}}\left(\sum_{l=0}^{\infty}\dfrac{(-1)^lz^l}{l!}\right) \left(\sum_{j=0}^{\infty} b_j^{(m)} z^j\right), \hsp \nonumber \\
&=& \dfrac{1}{z^{m+1}}\sum_{l=0}^{\infty}c_l^{(m)} z^l, \hsp \label{kbssl}
\end{eqnarray}
where
\begin{multline}
b_j^{(m)}=
\left\{
\begin{array}{l}
\dfrac{(2m-j)!}{j!(m-j)!2^{m-j}}, \hspace{2.5cm} \text{for} \quad j \le m,\\
0, \hspace{4.9cm} \text{for} \quad j>m.
\end{array}
\right. \hsp\label{bcoef}
\end{multline}
The coefficient $c_l^{(m)}$ is given by the Cauchy product,
\begin{equation}
c_l^{(m)} = \sum_{q=0}^{l}b_q^{(m)} \dfrac{(-1)^{l-q}}{(l-q)!} = \sum_{q=0}^{\min{(l,m)}} \dfrac{(2m-q)!}{(m-q)!q!2^{m-q}}\dfrac{(-1)^{l-q}}{(l-q)!}, \hsp \label{kcoef}
\end{equation}
where we have taken heed of the limited extent of $b_l^{(m)}$.

Consequently, we have the derivative, $k_m^{'}$:
\begin{equation}
k_m^{'}(z) = \sum_{l=0}^{\infty}c_l^{(m)}(l-m-1) z^{l-m-2} = \dfrac{1}{z^{m+2}}\sum_{l=0}^{\infty}c_l^{(m)}(l-m-1) z^l, \hsp \label{kbsslpr}
\end{equation}
The divergences of $k_m$ and $k_m^{'}$ thus stem from the factors of $1/z^m$ and $1/z^{m+1}$, respectively. The ascending series multiplying these factors, on the other hand are absolutely and uniformly convergent.

We now make use of these series expressions to develop expansions for products of Bessel functions or their derivatives, for different but related arguments, $x$ and $y$. Anticipating, ultimately, the development of small $s$ expansions, we shall develop series expansions in the variable $x$ with coefficients that will involve the ratio $y/x$. We begin with $k_m i_m^{'}$ and $k_m^{'}i_m$. For $ 0 < x,y \ll 1$,
\begin{eqnarray}
k_m(x) i_m^{'}(y) &=& \left(\dfrac{1}{x^{m+1}}\sum_{l=0}^{\infty}c_l^{(m)} x^l\right)\left(y^{m-1}\sum_{j=0}^{\infty} (2j+m) d_j^{(m)} y^{2j}\right), \nonumber\\
&=& \dfrac{y^{m-1}}{x^{m+1}}\left(\sum_{l=0}^{\infty}c_l^{(m)} x^l\right)\left(\sum_{j=0}^{\infty} e_j^{(m)} x^j\right) = \dfrac{y^{m-1}}{x^{m+1}}\sum_{l=0}^{\infty}g_l^{(m)} x^l, \hsp \label{kninpr}
\end{eqnarray}
where the explicit Cauchy product,
\begin{equation}
g_l^{(m)} = \sum_{q=0}^{l}e_q^{(m)} c_{l-q}^{(m)}, \hsp \label{gcoef}
\end{equation}
defines the new coefficient $g_l^{(m)}$, and where, in  \eqref{kninpr}, we have introduced,
\begin{multline}
e_q^{(m)}=
\left\{
\begin{array}{l}
(q+m) d_{q/2}^{(m)} \left(\dfrac{y}{x}\right)^{q} = \dfrac{(q+m)}{(q/2)!(2m+q+1)!!2^{q/2}}\left(\dfrac{y}{x}\right)^{q}, \quad \text{for} \quad q = 2j \text{ (even)},\\
0, \hspace{8.8cm} \text{for} \quad q =2j+1 \text{ (odd)},
\end{array}
\right. \hsp\label{ecoef}
\end{multline}
to convert the series in $y^2$ to one in $y$, and then one in $x$. In an analogous manner, for $ 0 < x,y \ll 1$ we obtain the expansion for $k_m^{'}i_m$,
\begin{equation}
k_m^{'}(x) i_m(y) = \dfrac{y^{m}}{x^{m+2}}\sum_{l=0}^{\infty}h_l^{(m)} x^l; \quad \quad \quad \quad h_l^{(m)} = \sum_{q=0}^{l}f_q^{(m)} (l-q-m-1) c_{l-q}^{(m)}, \hsp \label{knprin}
\end{equation}
with
\begin{multline}
f_q^{(m)}=
\left\{
\begin{array}{l}
d_{q/2}^{(m)} \left(\dfrac{y}{x}\right)^{q} = \dfrac{1}{(q/2)!(2m+q+1)!!2^{q/2}}\left(\dfrac{y}{x}\right)^{q}, \hspace{1.465cm} \text{for} \quad q = 2j \text{ (even)},\\
0, \hspace{8.6cm} \text{for} \quad q =2j+1 \text{ (odd)}.
\end{array}
\right. \hsp\label{fcoef}
\end{multline}
Similarly, for $ 0 < x,y \ll 1$, we have
\begin{equation}
i_m(x)i_m^{'}(y) = x^m y^{m-1} \sum_{j=0}^{\infty} t_j^{(m)} x^{2j} \quad \quad ; \quad \quad
i_m^{'}(x)i_m(y) = x^{m-1}y^m \sum_{j=0}^{\infty} T_j^{(m)} x^{2j}, \label{ininpr}
\end{equation}
with
\begin{equation}
t_j^{(m)} = \sum_{l=0}^{j} d_{j-l}^{(m)} d_{l}^{(m)} \left(2l+m\right)\left(\dfrac{y}{x}\right)^{2l} \quad \quad ; \quad \quad T_j^{(m)} = \sum_{l=0}^{j} d_l^{(m)} d_{j-l}^{(m)} \left(2l+m\right)\left(\dfrac{y}{x}\right)^{2(j-l)}. \hsp \label{tjTjcoef}
\end{equation}

Finally, we require an expansion of the product of two modified spherical Bessel functions of the first kind, $i_m (x)i_m (y)$. Again, for $ 0 < x,y \ll 1$,
\begin{equation}
i_m(x)i_m(y) = \left(\sum_{j=0}^{\infty} d_j^{(m)} x^{2j+m}\right)\left(\sum_{l=0}^{\infty} d_l^{(m)} y^{2l+m}\right) = \left(xy\right)^{m}\sum_{l=0}^{\infty} \Theta_l^{(m)} x^{2l}, \label{inin}
\end{equation}
with
\begin{equation}
\Theta_l^{(m)}(y,x) = \sum_{j=0}^{l} d_j^{(m)}d_{l-j}^{(m)}\left(\dfrac{y}{x}\right)^{2j}. \label{Thtacoef}
\end{equation}

\subsection{Expansions Specific to the Separation-of-Variable Solution}
Our focus in this section is to utilize the expansions given in the preceding section in order to develop Laplace transform parameter expansions of the separation of variable solution coefficients $\{A_m,B_m,E_m\}$ derived in Section \ref{sec:solution}. Our first goal is to develop expansions of the functions $\Lambda_m^{(1)}$ and $\Lambda_m^{(2)}$. To this end it proves convenient to first rewrite the expressions in  \eqref{lmda12} for the generic arguments $x$ and $y$:
\begin{equation}
\Lambda_m^{(1)} = \dfrac{1}{\beta}\left[\gamma k_m(x)i_m^{'}(y)-k_m^{'}(x)i_m(y)\right]; \quad \quad \quad
\Lambda_m^{(2)} = \dfrac{1}{\beta}\left[i_m^{'}(x)i_m(y)-\gamma i_m(x)i_m^{'}(y)\right]. \hsp \label{lmbda12_xy}
\end{equation}

Using expansions  \eqref{kninpr}, \eqref{knprin} and \eqref{ininpr} for the various products of the modified spherical Bessel functions $i_m$, $k_m$, $i_m^{'}$ and $k_m^{'}$, we readily deduce that for $0 < x,y \ll 1$,
\begin{equation}
\Lambda_m^{(1)} = \dfrac{1}{\beta}\dfrac{1}{x^2}\left(\dfrac{y}{x}\right)^{m-1}\sum_{l=0}^{\infty}\left[\gamma g_l^{(m)}-\dfrac{y}{x}h_l^{(m)}\right]x^l ; \quad \quad \quad
\Lambda_m^{(2)} = \dfrac{\left(xy\right)^{m-1}}{\beta}\sum_{l=0}^{\infty}\left[\dfrac{y}{x} T_l^{(m)}-\gamma t_l^{(m)}\right]x^{2l+1}.
\hsp \label{Lmda12exp1}
\end{equation}

\subsection{Arguments for the Separation-of-Variable Solution Expansions}
In the series expansions just developed we will consider arguments relevant to our problem. In the majority of cases, specifically  \eqref{ibsslpr}-\eqref{tjTjcoef}, the arguments we need to consider are as outlined in  \eqref{xygammayonx1} below. Note the key feature that both $x$ and $y$ variables are proportional to $\sqs$ and so we have (a) their ratio, and powers of their ratio, playing prominent roles in the coefficients of many of the expansions, and (b) their ratio being independent of $\sqs$.
\begin{multline}
\left\{
\begin{array}{l}
x = \sqrt{a^2/D_{I}}\sqs, \\
y = \sqrt{(a-\upsilon)^2/D_{II}} \sqrt{s} \doteq \sqrt{\dfrac{a^2}{D_{I}}}\sqrt{\dfrac{D_{I}}{D_{II}}}\left(1-\epsilon\right)\sqrt{s},\\
\gamma = \dfrac{\beta}{\alpha}\sqrt{\dfrac{D_{II}}{D_I}},
\end{array}
\right.
\quad \Rightarrow \quad
\left\{
\begin{array}{l}
\dfrac{y}{x} = \sqrt{\dfrac{D_{I}}{D_{II}}}\left(1-\epsilon\right), \\
xy = s\dfrac{a^2}{D_I} \left(1-\epsilon\right)\sqrt{\dfrac{D_I}{D_{II}}}. \\
\end{array}
\right. \hsp \label{xygammayonx1}
\end{multline}
Note that apart from $y/x$, $\gamma$ is also independent of the Laplace parameter, while $xy$ is proportional to $s$. In the above we have introduced the non-dimensional parameter $\epsilon = \upsilon/a \ll 1$.

With the above expressions as arguments of the Bessel functions, the coefficients in the square brackets of the two series in  \eqref{Lmda12exp1} are independent of the Laplace parameter $s$; the dependence on $s$ is contained either in the additional factors multiplying each series, or in the powers of $x \sim \sqs $ that make up the two series.

In the cases of  \eqref{inin}-\eqref{Thtacoef}, and equations derived from them, we employ different $x$ and $y$ arguments, namely,
\begin{multline}
\left\{
\begin{array}{l}
x = \sqrt{R^2/D_{I}}\sqs, \quad R \gg a, \\
y = \sqrt{r^2/D_{II}} \sqrt{s} \quad \doteq \dfrac{r}{R}\sqrt{\dfrac{D_{I}}{D_{II}}}\left(\sqrt{R^2/D_{I}}\right)\sqrt{s},
\end{array}
\right.
\quad \Rightarrow \quad
\left\{
\begin{array}{l}
\dfrac{y}{x} = \dfrac{r}{R} \sqrt{\dfrac{D_{I}}{D_{II}}}, \\
\\
xy = s\dfrac{rR}{\sqrt{D_I D_{II}}}, \\
\end{array}
\right. \hsp \label{xygammayonx2}
\end{multline}
where $0 \le r \le a - \upsilon$. Once again, both $x$ and $y$ here are proportional to $\sqs$, so that the ratio $y/x$ is independent of $s$.

\renewcommand{\theequation}{C.\arabic{equation}}
\section{Series Expansions in the Laplace Transform Parameter} \label{sec:seriesexp}
Our goal ultimately is to determine the time dependent development of concentration of solutes anywhere in the cell. To this end not only do we need the coefficient $E_m$, we need the product of $E_m$ with the Bessel function $i_m(\sqrt{r^2/D_{II}}\sqs)$.\footnote{The only exception, of course, being at the sphere centre, for which $r=0$ and therefore $i_m \doteq 1$.} Given that the inclusion of yet another Bessel function but with a new argument brings with it an $s$ dependence in addition to the $s$ dependence intrinsic to $E_m$ (appearing below), it proves useful to express all quantities explicitly in power series in the Laplace parameter. Consequently, in this section we continue the series development, but now in terms of $\sqs$. This task is conveniently executed as the majority of our expansions are already in forms suitable for this conversion. That is, every series can be re-expressed as series in powers of $\sqs$.

To begin with we have, from  \eqref{xygammayonx1}, $x=\sqrt{a^2/D_I}\sqs$ and $y/x =\sqrt{D_I/D_{II}}(1-\epsilon)$ from which we deduce the series expressions
\begin{equation}
\Lambda_m^{(1)} = \dfrac{1}{s} \sum_{l=0}^{\infty}\Delta_l^{(m)} \left(\sqs\right)^l ; \quad \quad \quad
\Lambda_m^{(2)} = \dfrac{1}{\sqs}\sum_{l=0}^{\infty}\Omega_l^{(m)}s^{l+m},
\hsp \label{Lmda12exp2}
\end{equation}
where $\Delta_l^{(m)}$ and $\Omega_l^{(m)}$ are given by,
\begin{equation}
\Delta_l^{(m)} = \dfrac{1}{\beta} \left(\sqrt{\dfrac{D_I}{D_{II}}}(1-\epsilon)\right)^{(m-1)}\left(\sqrt{\dfrac{a^2}{D_I}}\right)^{(l-2)}\sum_{q=0}^l\left(\gamma e_q^{(m)}-\sqrt{\dfrac{D_I}{D_{II}}}(1-\epsilon)f_q^{(m)}(l-q-m-1)\right)c_{l-q}^{(m)}, \label{Delta}
\end{equation}
and
\begin{eqnarray}
\Omega_l^{(m)} = &&\dfrac{1}{\beta}\left(\dfrac{a^2(1-\epsilon)}{D_I}\sqrt{\dfrac{D_I}{D_{II}}}\right)^{(m-1)} \dfrac{l}{2}\left(\dfrac{a^2}{D_I}\right)^{(2l+1)} \nonumber \\ &\times&\sum_{q=0}^ld_q^{(m)}d_{l-q}^{(m)}(2q+m)\left(\left(\sqrt{\dfrac{D_I}{D_{II}}}(1-\epsilon)\right)^{(2l-2q+1)} -\gamma\left(\sqrt{\dfrac{D_I}{D_{II}}}(1-\epsilon)\right)^{(2q)}\right). \nonumber \\
&&\label{Omega}
\end{eqnarray}
Secondly, present in the denominator of  \eqref{Emcoef} are modified spherical Bessel functions with argument $x=\sqrt{R^2/D_I}\sqs$. Substituting this new argument into  \eqref{ibsslpr} and \eqref{kbssl} we have, respectively,
\begin{equation}
i_m\left(\sqrt{\dfrac{R^2}{D_I}}\sqs\right) = \sum_{j=0}^{\infty} d_j^{(m)} \left(\dfrac{R^2}{D_{I}}\right)^{j+m/2} s^{j+m/2} \quad ; \quad k_m\left(\sqrt{\dfrac{R^2}{D_I}}\sqs\right) = \sum_{j=0}^{\infty} c_j^{(m)} \left(\sqrt{\dfrac{R^2}{D_I}}\right)^{j-(m+1)} \left(\sqs\right)^{j-(m+1)}. \label{bsslSeries}
\end{equation}
Multiplying the respective factors in  \eqref{Lmda12exp2} and \eqref{bsslSeries} we get, after some labour, the denominator of  \eqref{Emcoef} developed in a power series in $\sqs$, namely,
\begin{equation}
i_m\left(\sqrt{R^2/D_I}\sqs\right)\Lambda_m^{(1)} + k_m\left(\sqrt{R^2/D_I}\sqs\right)\Lambda_m^{(2)} = s^{m/2-1}\sum_{q=0}^{\infty} \left(\mathbb{X}_q^{(m)} + \mathbb{Y}_q^{(m)} \right) \left(\sqs\right)^q. \label{denom}
\end{equation}
Note that the two terms on the left hand side of the equation contribute to the same order of $\sqs$. In  \eqref{denom}, the $s$-independent coefficients, $\mathbb{X}_q^{(m)}$ and $\mathbb{Y}_q^{(m)}$, are given as Cauchy products of coefficients of the ingoing series:
\begin{equation}
\mathbb{X}_q^{(m)} = \sum_{p=0}^q \widetilde{\Phi}_p^{(m)} \Delta_{q-p}^{(m)} \quad ; \quad \mathbb{Y}_q^{(m)} = \sum_{p=0}^q \Psi_{q-p}^{(m)}\widetilde{\Omega}_p^{(m)}, \label{XmYm}
\end{equation}
with
\begin{equation}
\Psi_{q-p}^{(m)} = c_{q-p}^{(m)}\left(\sqrt{\dfrac{R^2}{D_I}}\right)^{(q-p-(m+1))},  \hsp \label{Phi}
\end{equation}
and
\begin{multline}
\widetilde{\Phi}_p^{(m)}=
\left\{
\begin{array}{l}
d_{p/2}^{(m)} \left(\dfrac{R^2}{D_{I}}\right)^{p/2+m/2} \hspace{-1cm}, \hspace{1cm}\\
0,
\end{array}
\right. \quad \quad ; \quad \quad
\widetilde{\Omega}_p^{(m)}=
\left\{
\begin{array}{l}
\Omega_{p/2}^{(m)}, \hspace{1.7cm} \quad \text{for} \quad p = 2j \text{ (even)}, \hsp\\
0, \hspace{2.7cm} \text{for} \quad p =2j+1 \text{ (odd)}. \hsp
\end{array}
\right.
\hsp\label{tldePhiOmega}
\end{multline}
Extracting the first and therefore dominant, $s$-independent term, $\left(\mathbb{X}_0^{(m)}+\mathbb{Y}_0^{(m)}\right)$, the denominator of  \eqref{Emcoef},  \eqref{denom}, takes the form
\begin{equation}
s^{m/2-1}\left(\mathbb{X}_0^{(m)}+\mathbb{Y}_0^{(m)}\right)\left(1+\mathrm{Z}^{(m)}\right), \label{denom2}
\end{equation}
where we have introduced $\mathrm{Z}^{(m)}$ for the $\sqs$-series,
\begin{equation}
\mathrm{Z}^{(m)}=\sum_{q=1}^{\infty}\dfrac{\left(\mathbb{X}_q^{(m)}+\mathbb{Y}_q^{(m)}\right)}{\left(\mathbb{X}_0^{(m)}+\mathbb{Y}_0^{(m)}\right)} \left(\sqs\right)^q \doteq \sum_{q=1}^{\infty}\dfrac{\mathbb{Z}_q^{(m)}}{\mathbb{Z}_0^{(m)}} \left(\sqs\right)^q \doteq \sum_{q=1}^{\infty}\mathcal{Z}_q^{(m)}\left(\sqs\right)^q. \label{Zeqn}
\end{equation}
The factor $\left(1+\mathrm{Z}^{(m)}\right)$ in  \eqref{denom2}, which contributes to the denominator of  \eqref{Emcoef}, with $\mathrm{Z}^{(m)} = O\left(\sqs\right)$, can be reconsidered as a factor $1/\left(1+\mathrm{Z}^{(m)}\right)$ \emph{multiplying the numerator} of  \eqref{Emcoef} and can then, for $\sqs \ll 1$, be re-expressed formally in the series
\begin{equation}
\dfrac{1}{1+\mathrm{Z}^{(m)}} = \sum_{l=0}^{\infty} (-1)^l \left(\mathrm{Z}^{(m)}\right)^l = \sum_{l=0}^{\infty} \mathbb{Q}_l^{(m)} \left(\sqs\right)^l,
\end{equation}
where the first few coefficients of the latter series are
\begin{eqnarray}
\mathbb{Q}_0^{(m)}&=&1, \nonumber \\
\mathbb{Q}_1^{(m)}&=&-\mathcal{Z}_1^{(m)}, \nonumber \\ \mathbb{Q}_2^{(m)}&=&-\mathcal{Z}_2^{(m)}+\left(\mathcal{Z}_1^{(m)}\right)^2, \label{Qcoef} \\ \mathbb{Q}_3^{(m)}&=&-\mathcal{Z}_3^{(m)}+2\mathcal{Z}_1^{(m)}\mathcal{Z}_2^{(m)}-\left(\mathcal{Z}_1^{(m)}\right)^3, \text{ and} \nonumber\\
\mathbb{Q}_4^{(m)} &=& -\mathcal{Z}_4^{(m)}+\left(\mathcal{Z}_2^{(m)}\right)^2 + 2\mathcal{Z}_1^{(m)}\mathcal{Z}_3^{(m)} + 3\left(\mathcal{Z}_1^{(m)}\right)^2\mathcal{Z}_2^{(m)} +\left(\mathcal{Z}_1^{(m)}\right)^4. \nonumber
\end{eqnarray}

Given all the preceding results we find that the interior solution coefficients, $E_m$, multiplied by the modified spherical Bessel function $i_m\left(\sqrt{r^2/D_{II}}\sqs\right)$, can similarly be expressed in series form. As there are many factors involved in arriving at the final result(s), we shall go through the argument in some detail, starting with the formal separation of variables expression, in so much as it involves the Laplace parameter.
\begin{eqnarray}
E_m i_m\left(\sqrt{r^2/D_{II}}\sqs\right) & \doteq & \dfrac{D_{I}}{s a^2} \dfrac{\dfrac{(2m+1)}{2}\Gamma_m}{\left[i_m\left(\sqrt{R^2/D_{I}}\sqs\right)\Lambda_m^{(1)}+k_m\left(\sqrt{R^2/D_{I}}\sqs\right)\Lambda_m^{(2)}\right]}i_m\left(\sqrt{r^2/D_{II}}\sqs\right), \nonumber \\
\nonumber \\
&=& \dfrac{D_{I}}{s a^2}\dfrac{\dfrac{(2m+1)}{2}\dfrac{2C_0}{s}\e^{-\left(\sqrt{z_0^2/D_{I}}\sqs\right)} i_m\left(\sqrt{R^2/D_{I}}\sqs\right)}{s^{(m/2-1)}\left(\mathbb{X}_0^{(m)}+\mathbb{Y}_0^{(m)}\right)\left(1+\mathrm{Z}^{(m)}\right)}
i_m\left(\sqrt{r^2/D_{II}}\sqs\right) \nonumber \\
\nonumber \\
&=& C_0\dfrac{D_{I}}{a^2}\dfrac{(2m+1)}{s^{(m/2+1)}\mathbb{Z}_0^{(m)}}\e^{-\left(\sqrt{z_0^2/D_{I}}\sqs\right)} i_m\left(\sqrt{R^2/D_{I}}\sqs\right)i_m\left(\sqrt{r^2/D_{II}}\sqs\right)\dfrac{1}{\left(1+\mathrm{Z}^{(m)}\right)}
\hsp \nonumber \\
\nonumber \\
&=& C_0\dfrac{D_{I}}{a^2}\dfrac{(2m+1)}{s^{(m/2+1)}\mathbb{Z}_0^{(m)}} \e^{-\left(\sqrt{z_0^2/D_{I}}\sqs\right)} \nonumber \\
&& \quad \quad \quad \quad \quad \times \left(s^m\left(\dfrac{rR}{\sqrt{D_{I}D_{II}}}\right)^{m}\sum_{l=0}^{\infty} \Theta_l^{(m)}(r/\sqrt{D_{II}},R/\sqrt{D_{I}}) \left(R^2/D_{I}\right)^l s^l \right) \nonumber \\
\nonumber \\
&& \hspace{8cm} \times \left(\sum_{q=0}^{\infty} (-1)^q \left(\mathrm{Z}^{(m)}\right)^q  \right), \nonumber \\
&=& C_0\dfrac{D_{I}}{a^2}\dfrac{(2m+1)}{\mathbb{Z}_0^{(m)}} \left(\dfrac{rR}{\sqrt{D_{I}D_{II}}}\right)^{m} s^{(m/2-1)}~\e^{-\left(\sqrt{z_0^2/D_{I}}\sqs\right)}\sum_{q=0}^{\infty} \mathbb{T}_q^{(m)}\left(\sqs\right)^q, \label{interim} \\
\nonumber \\
&=& C_0\dfrac{D_{I}}{a^2}\dfrac{(2m+1)}{\mathbb{Z}_0^{(m)}}\left(\dfrac{rR}{\sqrt{D_{I}D_{II}}}\right)^{m}s^{(m/2-1)}~\sum_{q=0}^{\infty} \mathbb{S}_q^{(m)}\left(\sqs\right)^q, \hsp \label{Emcoefs}
\end{eqnarray}
where
\begin{equation}
\mathbb{T}_q^{(m)} = \sum_{p=0}^q  \widetilde{\Theta}_p^{(m)}\mathbb{Q}_{q-p}^{(m)} \quad \text{and} \quad \mathbb{S}_q^{(m)} = \sum_{p=0}^q (-1)^p \dfrac{\left(\sqrt{z_0^2/D_{I}}\right)^{p}}{p!} \mathbb{T}_{q-p}^{(m)}, \hsp \label{SandT}
\end{equation}
and where we have introduced,
\begin{multline}
\widetilde{\Theta}_p^{(m)}=
\left\{
\begin{array}{l}
\Theta_{p/2}^{(m)}(r/\sqrt{D_{II}},R/\sqrt{D_{I}})\left(R^2/D_{I}\right)^{p/2}, \hspace{2cm} \quad \text{for} \quad p = 2l \text{ (even)}, \hsp\\
0, \hspace{8.2cm} \text{for} \quad p =2l+1 \text{ (odd)}, \hsp
\end{array}
\right.
\hsp\label{tldeTheta}
\end{multline}
to take care of a conversion from a series in $s$ to one in $\sqs$.

Since we are to take the inverse Laplace transform to return to the temporal domain one may use either of \eqref{interim} or \eqref{Emcoefs}, since the inverse transform of both are known. Equation \eqref{interim} has the advantage that it improves on the convergence for small $s$, and subsequently for large time.

\renewcommand{\theequation}{D.\arabic{equation}}
\section{Laplace Inversion of the Series Expansions}\label{sec:lapinvser}
In order to return to the temporal domain, the solution in the Laplace transformed domain needs to be inverted. This is achieved through evaluating the complex line integral,
\begin{equation}
c_{\varsigma}(r,\phi,t) = \mathfrak{L}^{-1}\left\{\overline{c}_{\varsigma}\right\} = \dfrac{1}{2 \pi i} \int_{\sigma - i \infty}^{\sigma + i \infty} \overline{c}_{\varsigma}(r,\phi,s) \e^{st} \d s, \quad \quad \varsigma = I,II, \label{LplcInv}
\end{equation}
which may be achieved either by resorting to a table of known transforms, or by performing the integration directly. As is usual, the real-valued constant, $\sigma$ is taken large enough to place the Bromwich line contour to the right of all singularities of $\overline{c}_{\varsigma}$~\citep{arfken2001,mf1953}. With the radius-dependent coefficients $E_m \times i_m\left(\sqrt{r^2/D_{II}}\sqs\right)$ each determined as an infinite series in $\sqs$, the inversion will also summarily be in the form of an infinite series in $t$. Since we consider two alternative expressions for the coefficients, one retaining the exponential $\exp\left(-\sqrt{z_0^2/D_{I}}\sqs\right)$ and one incorporating the exponential, we shall likewise obtain two corresponding, time-dependent series solutions.

We consider first the form where the exponential factor $\exp\left(-\sqrt{z_0^2/D_{I}}\sqs\right)$ is retained. From  \eqref{interim} we see that it is necessary to consider the Laplace inverse of terms in a double series, one series indexed by $m$, the order of the Legendre polynomial, and the other by $q$, the order of the perturbation series. To be precise, the inverse Laplace transform is of terms involving the exponential multiplied by $\sqs$ raised to an exponent in the combination $m+q$. We therefore need to consider a development that is conditional on the value of $m+q$. If $m+q$ is an even integer we consider Laplace inversions of the form
\begin{equation}
\hsp \mathfrak{L}^{-1}\left\{\e^{-b \sqs} s^{(L-1)}\right\}, \hspace{4cm}  m+q=2L, \quad L=0,1,2,\ldots~. \label{LinvExpEven}
\end{equation}
Alternatively, for odd integer values of $m+q$, we consider Laplace inversions,
\begin{equation}
\hsp \mathfrak{L}^{-1}\left\{\e^{-b \sqs} s^{(L-1/2)}\right\}, \hspace{4.2cm}  m+q=2L+1, \quad L=0,1,2,\ldots~. \label{LinvExpOdd}
\end{equation}
The following results may readily be verified,
\begin{multline}
\left\{
\begin{array}{l}
\mathfrak{L}^{-1}\left\{\dfrac{\e^{-\sqrt{z_0^2/D_{I}} \sqs}}{s}\right\} = 1-\text{erf}\left(\dfrac{z_0}{\sqrt{4 D_I t}}\right), \hspace{5.5cm} m+q=0, \\
\mathfrak{L}^{-1}\left\{\dfrac{\e^{-\sqrt{z_0^2/D_{I}} \sqs}}{\sqs}\right\} = \dfrac{1}{\sqrt{\pi}} \e^{\left(-\dfrac{z_0^2}{4 D_I t}\right)}\dfrac{1}{\sqrt{t}}, \hspace{5.55cm} m+q=1, \\
\mathfrak{L}^{-1}\left\{\e^{-\sqrt{z_0^2/D_{I}} \sqs}\right\} = \dfrac{z_0}{\sqrt{4\pi D_I}}\e^{\left(-\dfrac{z_0^2}{4 D_I t}\right)}\dfrac{1}{t^{3/2}}, \hspace{4.9cm} m+q=2, \\
\mathfrak{L}^{-1}\left\{\e^{-\sqrt{z_0^2/D_{I}} \sqs} \cdot \sqs\right\} = \dfrac{1}{\sqrt{4\pi}}\e^{\left(-\dfrac{z_0^2}{4 D_I t}\right)} \left(\dfrac{z_0^2}{4 D_I t^{5/2}}-\dfrac{1}{t^{3/2}}\right), \hspace{2.3cm} m+q=3, \\
\mathfrak{L}^{-1}\left\{\e^{-\sqrt{z_0^2/D_{I}} \sqs} \cdot s \right\} = \dfrac{z_0}{\sqrt{4 \pi D_I}}\e^{\left(-\dfrac{z_0^2}{4 D_I t}\right)}\left(\dfrac{z_0^2}{8 D_I t^{7/2}}-\dfrac{1}{t^{5/2}}\right), \hspace{2.1cm} m+q=4.
\end{array}
\right. \hsp \label{FstOrdInv1}
\end{multline}
Higher order contributions may readily be determined by successive differentiation. We see in  \eqref{FstOrdInv1}, a concentration profile in the cell interior which comprises terms that are a mix of an asymptotic solution in negative powers of $t^{1/2}$ and an exponential decay with $t^{-1}$.

In contrast, by incorporating the remaining exponential into a single series in $\sqs$, as in  \eqref{Emcoefs}, we are left to consider inverse Laplace transforms of the form,
\begin{equation}
\hsp \mathfrak{L}^{-1}\left\{s^{(L-1)}\right\}, \hspace{4cm} m+q=2L, \quad L=0,1,2,\ldots, \label{LinvSEven}
\end{equation}
or,
\begin{equation}
\hsp \mathfrak{L}^{-1}\left\{s^{(L-1/2)}\right\}, \hspace{4.2cm} m+q=2L+1, \quad L=0,1,2,\ldots, \label{LinvSOdd}
\end{equation}
according to the evenness or oddness of $m+q$. The calculations of  \eqref{LinvSEven} and \eqref{LinvSOdd} are well known. In fact, the inversion of \eqref{LinvSEven}, when $m+q=2L$ results in
\begin{multline}
\hsp \mathfrak{L}^{-1}\left\{s^{(L-1)}\right\} =
\left\{
\begin{array}{l}
1, \hspace{3cm} m+q=0, \hsp \\
0, \hspace{3cm} m+q=2L, \quad L =1,2,3, \hsp \ldots~.
\end{array}
\right. \hsp \label{FstOrdInv2a}
\end{multline}
When $m+q=2L$, with $L \ge 1$, the argument is an analytic function with no singularities in the complex plane. Hence the integral is zero by Cauchy's theorem \citep{arfken2001,mf1953}. For the case $m+q=0$, the inverse is unity.

In contrast, when $m+q=2L+1$ there is again a branch point at the origin, it may then readily be determined by explicit calculation of the Bromwich contour integral~\citep{arfken2001,mf1953} or resorting to a reliable table of transforms \citep{abram1965}, that
\begin{equation}
\mathfrak{L}^{-1}\left\{s^{(L-\frac{1}{2})}\right\} = \dfrac{1}{\sqrt{\pi}} \dfrac{\partial^L}{\partial t^L} \left(\dfrac{1}{t^{1/2}}\right), \hspace{3cm} m+q=2L+1, \quad L=0,1,2, \ldots~, \hsp \label{FstOrdInv2b}
\end{equation}
which amounts to a concentration profile in the cell interior that is given strictly as an asymptotic series in inverse powers of $t^{1/2}$, with coefficients that depend on radius $r$ as well as other system parameters.

\renewcommand{\theequation}{E.\arabic{equation}}
\section{Large-time, Asymptotic Cell Concentration}\label{sec:largetimesolution}
Given the complicated nature of the solution, there are relevant reasons (to be explained as required) to consider a graded presentation of the solution. In the following subsections, we discuss first the steady state solution, proceeding then to an angular dependent approximation valid for ``extreme'' time values. We then consider the latter's orthogonal complement of a full time dependence at the expense of any angle dependence. Finally, the complete, formal asymptotic solution involving an angle-dependence valid over a larger time range is discussed. It is significant to point out that such a progressive discussion was not necessary in the works of \citep{philip1964} and \citep{mild1971} as they considered much simpler, spherically symmetric problems.

\subsection{Steady State Limit} \label{sec:stdystte} 
One means of estimating the accuracy of the asymptotic perturbation approach is to extract the steady state solution to see if it produces the expected property. This is done by considering only those terms in  \eqref{TrfAsymSln1} or \eqref{TrfAsymSln2} that are either independent of the Laplace parameter, or otherwise lead to a time-independent result.

Since it is clear that higher order terms in the $\sqs$ series will give rise to a time dependence, the case of $m+q=0$ is singled out for inspection. From \eqref{interior} the steady state result is then,
\begin{equation}
c_{II}(r, \phi, t \rightarrow \infty) = \lim_{t \rightarrow \infty} \mathfrak{L}^{-1}\left\{E_0 i_0\left(\dfrac{r}{\sqrt{D_{II}}}\sqs \right)\right\}. \hsp \label{stdystte}
\end{equation}
In the case of  \eqref{TrfAsymSln1_y} in which the exponential $\e^{-\left(\sqrt{z_0^2/D_{I}}\sqs\right)}$ has been incorporated into the series in $\sqs$, the inversion of the $m+q=0$ term leads to (see Appendix D,  \eqref{FstOrdInv2a})
\begin{equation}
c_{II}(r, \phi, t \rightarrow \infty) = \lim_{t \rightarrow \infty} \left\{C_0 \dfrac{D_{I}}{a^2}\dfrac{\mathbb{S}_0^{(0)}}{\mathbb{Z}_0^{(0)}}\cdot1\right\} = \left\{C_0 \dfrac{D_{I}}{a^2}\dfrac{\mathbb{S}_0^{(0)}}{\left(\mathbb{X}_0^{(0)}+\mathbb{Y}_0^{(0)}\right)}\right\}. \hsp \label{stdystte1}
\end{equation}
From formulae  \eqref{ibsslpr}, \eqref{Thtacoef}, \eqref{Delta}, \eqref{Phi}, \eqref{tldePhiOmega}, \eqref{Zeqn}, \eqref{Qcoef}, and \eqref{SandT}, given in Appendices B-C, it may readily be verified that
\begin{multline}
\left\{
\begin{array}{l}
\mathbb{S}_0^{(0)} = \mathbb{T}_0^{(0)} = \Theta_0^{(0)} \mathbb{Q}_0^{(0)} = \left(d_0^{(0)}\right)^2 = 1; \hsp \\
\mathbb{Z}_0^{(0)}=\mathbb{X}_0^{(0)}+\mathbb{Y}_0^{(0)} = \widetilde{\Phi}_0^{(0)} \Delta_0^{(0)} + \Psi_0^{(0)}\widetilde{\Omega}_0^{(0)}. \hsp
\end{array}
\right. \hsp \label{SandT00}
\end{multline}
In the last expression $\widetilde{\Phi}_0^{(0)} = d_0^{(0)} = 1$, while $\widetilde{\Omega}_0^{(0)}$ contains the factor $(2q+m)$ and is therefore zero for $q=m=0$. The remaining contributing factor to $\mathbb{Z}_0^{(0)}$ is ($\epsilon = v/a$)
\begin{equation}
\Delta_0^{(0)} = \dfrac{1}{\beta} \left(\sqrt{\dfrac{D_I}{D_{II}}}(1-\epsilon)\right)^{(-1)}\left(\sqrt{\dfrac{a^2}{D_I}}\right)^{(-2)}\left(\gamma e_0^{(0)}-\sqrt{\dfrac{D_I}{D_{II}}}(1-\epsilon)f_0^{(0)}(-1)\right)c_{0}^{(0)}. \label{Delta00}
\end{equation}
From inspection of  \eqref{kcoef}, \eqref{ecoef} and \eqref{fcoef}, we find $c_{0}^{(0)}=1$, $e_0^{(0)}=0$ and $f_0^{(0)}=1$. Consequently, $\mathbb{Z}_0^{(0)} = \Delta_0^{(0)} = \dfrac{1}{\beta}\dfrac{D_I}{a^2}$, and so
\begin{equation}
c_{II}(r, \phi, t \rightarrow \infty) = \left\{C_0 \dfrac{D_{I}}{a^2}\left(\dfrac{1}{\beta}\dfrac{D_I}{a^2}\right)^{-1}\right\} = \beta C_0,
\end{equation}
which is the expected steady state result.

Alternatively, considering the more rapidly convergent version of the asymptotic solution, in which the exponential $\e^{-\left(\sqrt{z_0^2/D_{I}}\sqs\right)}$ is retained, the Laplace inversion results in the contribution (Appendix D,  \eqref{FstOrdInv1})
\begin{equation}
c_{II}(r, \phi, t \rightarrow \infty) = \lim_{t \rightarrow \infty} \left\{C_0 \dfrac{D_{I}}{a^2}\dfrac{\mathbb{T}_0^{(0)}}{\mathbb{Z}_0^{(0)}}\cdot\left(1-\text{erf}\left(\dfrac{z_0}{\sqrt{4D_It}}\right)\right)\right\}, \hsp \label{stdystte2}
\end{equation}
which involves the same parameter values as  \eqref{SandT00} and \eqref{Delta00} but which possesses a residual time dependence. Nevertheless, in the limit $t \rightarrow \infty$ we find
\begin{equation}
c_{II}(r, \phi, t \rightarrow \infty) = \lim_{t \rightarrow \infty} \left\{\beta C_0 \cdot\left(1-\text{erf}\left(\dfrac{z_0}{\sqrt{4D_It}}\right)\right)\right\} = \beta C_0. \hsp \label{stdystte2x}
\end{equation}
That is, we deduce the same steady state limit of $\beta C_0$. The fact that some time dependence is present at this order of contribution reflects the fact that retaining the exponential improves convergence. It is interesting that at this order of approximation, the time dependence implies a gradual but \emph{uniform} build-up of particle concentration inside the cell due to the progressive rise in concentration in the surrounding environment.

Given the obviously superior performance of  \eqref{TrfAsymSln2_y} and in the interest of conserving space, we shall hereafter only quote those results relevant to this approximation, and simply comment where required on the results corresponding to the slower convergent series.

\subsection{Extreme Time, Angle Dependence} \label{sec:qeq0mGE0} 
It may be argued that the somewhat artificial, but the next simplest, approximation is to select the angle-dependent series with minimal contribution from the $\sqs$ series. This is achieved by encompassing all terms in the Legendre expansion ($m \ge 0$) while taking only the constant term ($q=0$) of the $\sqs$ series in  \eqref{TrfAsymSln1} and \eqref{TrfAsymSln2}. Under the premise that small $s$ is consonant with large $t$, other things being equal, neglecting all higher-order terms of the series is akin to reducing the range of validity to larger values of time. It therefore seems appropriate to refer to this approximation as an extreme time approximation. Despite all-but neglecting the $\sqs$-series there remains a time dependence in the final solution arising from the $s$-dependent factors multiplying the $\sqs$ series. It is perhaps worth mentioning that the necessity of \emph{some} time dependence consequential to an assumed angular dependence can be readily appreciated on physical grounds.

The advantage of singling out this approximation, which is also the objective here, is that it is one of the simpler cases with which to explicitly extract the physical parameter dependence of the respective terms and their coefficients. Under the stated conditions, the solution for the particle distribution inside the cell takes the form
\begin{equation}
c_{II}(r, \phi, t) \approx C_0\dfrac{D_{I}}{a^2}\sum_{m=0}^{\infty} \dfrac{(2m+1)}{\mathbb{Z}_0^{(m)}}\left(\dfrac{rR}{\sqrt{D_{I}D_{II}}}\right)^{m} \mathbb{T}_0^{(m)}~\mathfrak{L}^{-1}\left\{s^{(m/2-1)}\e^{-\left(\sqrt{z_0^2/D_{I}}\sqs\right)}\right\} P_m(\cos \phi), \hsp \label{interior_qeq0mGE0_2}
\end{equation}
The case $m=0$ is that discussed in the preceding subsection. The terms beyond the steady-state limit become more and more complex with increasing $m$, so we limit our discussion to the first few terms only. Invoking the inverse Laplace transform results in \eqref{FstOrdInv1} we obtain,
\begin{eqnarray}
c_{II}(r, \phi, t)
&\approx& C_0\dfrac{D_{I}}{a^2}\dfrac{1}{\sqrt{\pi}}\left\{ \sqrt{\pi}\dfrac{\mathbb{T}_0^{(0)}}{\mathbb{Z}_0^{(0)}}\left(1-\text{erf}\left(\dfrac{z_0}{\sqrt{4D_It}}\right)\right) +\dfrac{\e^{\left(-z_0^2/4D_It\right)}}{t^{1/2}}\left[ 3\left(\dfrac{rR}{\sqrt{D_ID_{II}}}\right)\dfrac{\mathbb{T}_0^{(1)}}{\mathbb{Z}_0^{(1)}}\cos \phi \right. \right.\nonumber \\
\nonumber \\
&&\hspace{2cm}+\dfrac{z_0}{\sqrt{4D_I}}\dfrac{1}{t}\dfrac{5}{2}\left(\dfrac{rR}{\sqrt{D_ID_{II}}}\right)^2\dfrac{\mathbb{T}_0^{(2)}}{\mathbb{Z}_0^{(2)}}(3\cos^2 \phi-1) \nonumber \\
\nonumber \\
&& \hspace{1.5cm}+\left. \left. \dfrac{1}{2t}\left(\dfrac{z_0^2}{4D_I t}-1\right) \dfrac{7}{2}\left(\dfrac{rR}{\sqrt{D_ID_{II}}}\right)^3\dfrac{\mathbb{T}_0^{(3)}}{\mathbb{Z}_0^{(3)}}(5\cos^3 \phi-3 \cos \phi) + \cdots \right] \right\}, \nonumber \\ \nonumber \\ \label{interior_qeq0mGE0_2x}
\end{eqnarray}
where, using  \eqref{ibsslpr}, \eqref{Thtacoef}, \eqref{Delta}, \eqref{Phi}, \eqref{tldePhiOmega}, \eqref{Zeqn}, \eqref{Qcoef}, and \eqref{SandT}, given in Appendices B-C, we find the following reduction of the coefficients,
\begin{multline*}
\left\{
\begin{array}{l}
\mathbb{T}_0^{(1)} = \Theta_0^{(1)} \mathbb{Q}_0^{(1)} = \left(d_0^{(1)}\right)^2  \quad ; \quad \mathbb{T}_0^{(2)} = \Theta_0^{(2)} \mathbb{Q}_0^{(2)} = \left(d_0^{(2)}\right)^2  \quad ; \quad \mathbb{T}_0^{(3)} = \Theta_0^{(3)} \mathbb{Q}_0^{(3)} = \left(d_0^{(3)}\right)^2\hsp \\
\mathbb{Z}_0^{(m)}=\mathbb{X}_0^{(m)}+\mathbb{Y}_0^{(m)} = \widetilde{\Phi}_0^{(m)} \Delta_0^{(m)} + \Psi_0^{(m)}\widetilde{\Omega}_0^{(m)} = d_0^{(m)}\left(\dfrac{R^2}{D_I}\right)^{m/2} \Delta_0^{(m)} + c_0^{(m)}\left(\dfrac{\sqrt{D_I}}{R}\right)^{(m+1)}\Omega_0^{(m)} \hsp
\end{array}
\right. 
\end{multline*}
\begin{equation}
 \label{SandT_qeq0mGE0_1}
 \end{equation}
With the various parameter inserts, the denominator, $\mathbb{Z}_0^{(m)}$, for general $m>0$, can be written as ($\epsilon=v/a$),
\begin{eqnarray}
\mathbb{Z}_0^{(m)}&=&\dfrac{1}{\beta}\dfrac{D_I}{a^2}\left(d_0^{(m)}\right)^2c_0^{(m)}\left(\sqrt{\dfrac{D_I}{D_{II}}}(1-\epsilon)\right)^{(m-1)}
\left(\dfrac{R^2}{D_I}\right)^{(m/2)} \hsp \nonumber \\
&\times& \left\{m\dfrac{\beta}{\alpha}\sqrt{\dfrac{D_{II}}{D_{I}}}\left[1- \left(\dfrac{a}{R}\right)^{(2m+1)}\right]+\sqrt{\dfrac{D_{I}}{D_{II}}}(1-\epsilon)\left[(m+1)+ m\left(\dfrac{a}{R}\right)^{(2m+1)}\right]\right\}. \hsp \nonumber \\
\end{eqnarray}
We see from these last three results that the factors $\mathbb{T}_0^{(m)}=\left(d_0^{(m)}\right)^2$, $D_I/a^2$ and $\left(R/\sqrt{D_I}\right)^m$ cancel from  \eqref{interior_qeq0mGE0_2x}, and a factor of $\beta$ is common to all terms. Thus, we can summarize the solution as
\begin{eqnarray}
c_{II}(r, \phi, t) &\approx& \beta C_0~\left\{1-\text{erf}\left(\dfrac{z_0}{\sqrt{4D_It}}\right) + \dfrac{1}{\sqrt{\pi}}\e^{\left(-\dfrac{z_0^2}{4 D_I t}\right)}\left[\left(\dfrac{r}{\sqrt{D_{II}}}\right) \dfrac{3}{Z_0^{(1)}}\dfrac{\cos \phi}{t^{1/2}} \right. \right. \hsp \nonumber\\
&& \quad \quad \quad + \dfrac{5}{2}\left(\dfrac{r}{\sqrt{D_{II}}}\right)^{2} \dfrac{z_0}{2\sqrt{D_I}Z_0^{(2)}}\dfrac{\left(3 \cos^2 \phi - 1\right)}{t^{3/2}} \hsp \nonumber\\
&& \quad \quad \quad \left.\left. + \dfrac{7}{4}\left(\dfrac{r}{\sqrt{D_{II}}}\right)^{3} \dfrac{1}{Z_0^{(3)}}\dfrac{\left(5 \cos^3 \phi - 3 \cos \phi\right)}{t^{3/2}}\left(\dfrac{z_0^2}{4D_It}-1\right) + \ldots \right] \right\}.\hsp \nonumber \\
\label{interior_qeq0mGE0_2y}
\end{eqnarray}
where the simplified but explicitly parameter-dependent denominators, $Z_0^{(1)}$ and $Z_0^{(3)}$, can be deduced (using  \eqref{kcoef}) from
\begin{eqnarray}
Z_0^{(m)}&=&\dfrac{(2m)!}{2^m (m-1)!}\left(\sqrt{\dfrac{D_I}{D_{II}}}(1-\epsilon)\right)^{(m-1)}
\hsp \nonumber \\
&\times& \left\{\dfrac{\beta}{\alpha}\sqrt{\dfrac{D_{II}}{D_{I}}}\left[1- \left(\dfrac{a}{R}\right)^{(2m+1)}\right]+\sqrt{\dfrac{D_{I}}{D_{II}}}(1-\epsilon)\left[\dfrac{(m+1)}{m}+ \left(\dfrac{a}{R}\right)^{(2m+1)}\right]\right\}, \hsp \nonumber \\
&& \hspace{10cm} m=1,2,3,\ldots \hsp \label{Z_qeq0meq13_1}
\end{eqnarray}

Explicitly retaining the exponential function in \eqref{TrfAsymSln2_y} not only leads to a more accurate time dependence, it also introduces a more accurate angle dependence than can be obtained with the slower convergent series, \eqref{TrfAsymSln1_y}. The latter fact becomes evident when one recalls that, according to  \eqref{FstOrdInv2b}, the inverse Laplace transform of all even-order contributions in the series \eqref{TrfAsymSln1_y} are identically zero, since those terms constitute analytic functions. The terms that do appear, at that level of approximation, contribute in increasing powers of $t^{-1/2}$, but without the exponential factor. Although there is no physical basis for doing so, the reader can readily uncover precisely how the series approximation corresponding to \eqref{interior_qeq0mGE0_2y} for the slower convergent case would be, by simply setting $z_0$ identically to zero in  \eqref{interior_qeq0mGE0_2y}. As a further point of contrast between \eqref{interior_qeq0mGE0_2y} and its slower convergent counterpart is the explicit dependence on the location of the plane source at $z=z_0$ in \eqref{interior_qeq0mGE0_2y}, rather than (only) via the parameter $R$, which enters in boundary condition \eqref{OuterBC}.

It is interesting that while the constant $\beta$ is prominent in the leading coefficient, the flux constant $\alpha$ enters in a relatively discrete way through the denominators, $Z_0^{(m)}$. On the other hand, it is \emph{not} surprising that the $\alpha$'s appearance is coincident with the $r$-dependence in the approximation (a minor indication of the correctness of the analysis). It is also interesting that $R$ does not explicitly appear in \eqref{interior_qeq0mGE0_2y}.

\subsection{Large-time, Angle Independence} \label{sec:meq0qGE0} 
We see in Section \ref{sec:particleuptake1D} that it is justified to consider the specialized case of no angle dependence, \textit{i.e.}, $m=0$, but with contributions from the full $\sqs$ series. In addition, being a single infinite series it is the next simplest approximation to study, as it reveals a considerable amount of explicit dependence on system parameters. As in the previous subsection (and the one to follow) we shall only treat the faster convergent approximation wherein the exponential is retained in its original form. The alternative expansion can be readily obtained by simply setting $z_0$ to zero.
\begin{equation}
c_{II}(r, \phi, t) = C_0\dfrac{D_{I}}{a^2} \dfrac{(1)}{\mathbb{Z}_0^{(0)}} \sum_{q=0}^{\infty} \mathbb{T}_q^{(0)}~\mathfrak{L}^{-1}\left\{s^{(q/2-1)}\e^{-\left(\sqrt{z_0^2/D_{I}}\sqs\right)}\right\}. \hsp \label{interior_meq0qGE0_2}
\end{equation}
At this level of approximation, using the results of Appendix D, we deduce the formal expression,
\begin{eqnarray}
c_{II}(r, \phi, t)
&=& \beta C_0\dfrac{1}{\sqrt{\pi}}\left\{ \sqrt{\pi}\left(1-\text{erf}\left(\dfrac{z_0}{\sqrt{4D_It}}\right)\right) +\dfrac{\e^{\left(-z_0^2/4D_It\right)}}{t^{1/2}}\left[\mathbb{T}_1^{(0)}
\right. \right. \nonumber \\
\nonumber \\
&& \left. \left. +\dfrac{z_0}{\sqrt{4D_I}}\dfrac{1}{t}\mathbb{T}_2^{(0)} + \dfrac{1}{2t}\left(\dfrac{z_0^2}{4D_I t}-1\right)\mathbb{T}_3^{(0)} + \dfrac{z_0}{\sqrt{4D_I}}\dfrac{1}{t^2}\left(\dfrac{z_0^2}{8D_I t}-1\right)\mathbb{T}_4^{(0)}\cdots \right] \right\}, \nonumber \\ \label{interior_meq0qGE0_2x}
\end{eqnarray}
where $\mathbb{T}_1^{(0)}=\left(d_0^{(0)}\right)^2=1$,
\begin{multline}
\left\{
\begin{array}{l}
\mathbb{T}_2^{(0)} = \mathbb{Q}_2^{(0)} + \dfrac{1}{6}\dfrac{R^2}{D_I}\left(1+\dfrac{r^2}{R^2}\dfrac{D_I}{D_{II}}\right), \hsp \\
\mathbb{T}_3^{(0)} = \mathbb{Q}_3^{(0)} - \dfrac{1}{6}\dfrac{R^2}{D_I}\left(1+\dfrac{r^2}{R^2}\dfrac{D_I}{D_{II}}\right) \dfrac{\Delta_1^{(0)}}{\mathbb{Z}_0^{(0)}}, \hsp \\
\mathbb{T}_4^{(0)} = \mathbb{Q}_4^{(0)} + \dfrac{1}{6}\dfrac{R^2}{D_I}\left(1+\dfrac{r^2}{R^2}\dfrac{D_I}{D_{II}}\right) \dfrac{\left(\left(\Delta_1^{(0)}\right)^2-\mathbb{Z}_2^{(0)}\mathbb{Z}_0^{(0)}\right)}{\left(\mathbb{Z}_0^{(0)}\right)^2} \hsp \\
\hspace{5cm} + \left(\dfrac{R^2}{D_I}\right)^2\left[d_2^{(0)} + \dfrac{1}{36}\dfrac{r^2}{R^2}\dfrac{D_I}{D_{II}}
+d_2^{(0)} \left(\dfrac{r^2}{R^2}\dfrac{D_I}{D_{II}}\right)^2 \right],
\end{array}
\right. \hsp \label{BigT2-40}
\end{multline}
where, once again, $\mathbb{Z}_0^{(0)} = \dfrac{1}{\beta} \dfrac{\D_I}{a^2}$. Of those constants present, it is straightforward to deduce that $\mathbb{Z}_1^{(0)}=\Delta_1^{(0)}=0$, $d_1^{(0)}=1/6$, $d_2^{(0)}=1/120$, and ($\epsilon = v/a$)
\begin{equation}
\dfrac{\mathbb{Z}_2^{(0)}}{\mathbb{Z}_0^{(0)}} = \dfrac{1}{6}\dfrac{R^2}{D_I}\left\{1+\dfrac{a^2}{R^2}
\left(\dfrac{D_I}{D_{II}}(1-\epsilon)^2+2\dfrac{\beta}{\alpha}(1-\epsilon)-3\right)+2\dfrac{a^3}{R^3}\left(1-\dfrac{\beta}{\alpha}(1-\epsilon)\right)\right\} \hsp \label{Z20onZ00}
\end{equation}
While the remaining constants, $\mathbb{Q}_q^{(0)}$ for example, are necessary for a complete description of the local concentration at this level of approximation, we give them no further attention as they do not feature in Section \ref{sec:discussion1}. However, their values can be deduced, as required, from  \eqref{ibsslpr}, \eqref{Delta}, \eqref{Zeqn} and \eqref{Qcoef}. The importance of  \eqref{interior_meq0qGE0_2x}, especially in regard to the total content of solutes, which is the focus of Section \ref{sec:discussion1}, lies in its explicit dependence on $t$ and $r$.

\subsection{Large-time, Angle Dependence} \label{sec:qmGE0} 
The results of the preceding subsections of course tell incomplete stories. A more complete description recognizes the fact that terms of equal $m+q$ will contribute at the same level of time order. For example, the first time-dependent term in  \eqref{interior_qeq0mGE0_2y} must be complemented with the term $m=0$, $q=1$. This latter term, according to  \eqref{FstOrdInv2b} and as explicit in  \eqref{interior_meq0qGE0_2x}, will similarly decay in time as $\exp\left(-z_0^2/4D_It\right)t^{-1/2}$, but possesses no angle dependence. Similar considerations apply to higher order contributions.

The formal expression for the local concentration internal to the cell, showing (at least) the radial, angular and time dependence, is thus,
\begin{eqnarray}
c_{II}(r, \phi, t)
&=& C_0\dfrac{D_{I}}{a^2}\dfrac{1}{\sqrt{\pi}}\left\{{\cor \sqrt{\pi}\dfrac{\mathbb{T}_0^{(0)}}{\mathbb{Z}_0^{(0)}}\left(1-\text{erf}\left(\dfrac{z_0}{\sqrt{4D_It}}\right)\right)} +\dfrac{\e^{\left(-z_0^2/4D_It\right)}}{t^{1/2}}\left\{\left[{\cob \dfrac{\mathbb{T}_1^{(0)}}{\mathbb{Z}_0^{(0)}}}
+{\cor 3\left(\dfrac{rR}{\sqrt{D_ID_{II}}}\right)\dfrac{\mathbb{T}_0^{(1)}}{\mathbb{Z}_0^{(1)}}\cos \phi}\right] \right. \right.\nonumber \\
\nonumber \\
&+&\dfrac{z_0}{\sqrt{4D_I}}\dfrac{1}{t}\left[{\cob \dfrac{\mathbb{T}_2^{(0)}}{\mathbb{Z}_0^{(0)}}}
+ 3\left(\dfrac{rR}{\sqrt{D_ID_{II}}}\right)\dfrac{\mathbb{T}_1^{(1)}}{\mathbb{Z}_0^{(1)}}\cos \phi+ {\cor\dfrac{5}{2}\left(\dfrac{rR}{\sqrt{D_ID_{II}}}\right)^2\dfrac{\mathbb{T}_0^{(2)}}{\mathbb{Z}_0^{(2)}}(3\cos^2 \phi-1)}\right] \nonumber \\
\nonumber \\
&+& \dfrac{1}{2t}\left(\dfrac{z_0^2}{4D_I t}-1\right)\left[{\cob \dfrac{\mathbb{T}_3^{(0)}}{\mathbb{Z}_0^{(0)}}} +3\left(\dfrac{rR}{\sqrt{D_ID_{II}}}\right)\dfrac{\mathbb{T}_2^{(1)}}{\mathbb{Z}_0^{(1)}}\cos \phi \right. \nonumber \\
\nonumber \\
&+& \left. \left. \left. \dfrac{5}{2}\left(\dfrac{rR}{\sqrt{D_ID_{II}}}\right)^2\dfrac{\mathbb{T}_1^{(2)}}{\mathbb{Z}_0^{(2)}}(3\cos^2 \phi-1) + {\cor \dfrac{7}{2}\left(\dfrac{rR}{\sqrt{D_ID_{II}}}\right)^3\dfrac{\mathbb{T}_0^{(3)}}{\mathbb{Z}_0^{(3)}}(5\cos^3 \phi-3 \cos \phi)}\right] + \cdots \right\} \right\}, \nonumber \\ \nonumber \\ \label{interior_qmGE0_1x}
\end{eqnarray}
where we have identified in red the terms captured in  \eqref{interior_qeq0mGE0_2y} and in blue the terms captured in  \eqref{interior_meq0qGE0_2x}, so as to highlight the other terms that are also present at the same order of time\footnote{Text color online. The reference to terms in blue and red in  \eqref{interior_qmGE0_1x} identifies the first and last terms in the square brackets, respectively, at each order of time decay.}.

Although it is cumbersome to provide explicit expressions for all the coefficients of the above, complete asymptotic series, it should be appreciated that the computation of the hierarchy of factors contributing to the coefficients is straightforward to implement in a numerical code. We demonstrate this in a separate publication where we compare our asymptotic calculations with a full numerical solution of the original mathematical problem using the finite element method~\cite{Kumar2024}.

\nolinenumbers

\end{document}